\pdfoutput=1
\def\buildtechreport{}

\documentclass[sigconf,nonacm]{acmart}

\usepackage{pvldb}

\renewcommand\vldbdoi{XX.XX/XXX.XX}
\renewcommand\vldbpages{XXX-XXX}
\renewcommand\vldbavailabilityurl{}

\usepackage{algorithm}
\usepackage{algpseudocode}
\usepackage{amsmath}
\usepackage{mathtools}
\usepackage{booktabs}
\usepackage{xcolor}

\newif\ifshowchanges
\showchangesfalse
\ifshowchanges\else\definecolor{blue}{rgb}{0,0,0}\fi

\newif\iftechreport
\ifdefined\buildtechreport\techreporttrue\fi
\iftechreport\renewcommand\vldbtopmatter{\pagestyle{\vldbpagestyle}}\fi
\newif\ifshowchanges
\showchangesfalse
\ifshowchanges\else\definecolor{blue}{rgb}{0,0,0}\fi
\usepackage{enumitem}
\usepackage{float}
\usepackage{wrapfig}

\microtypesetup{expansion=false}
\newtheorem{theorem}{Theorem}
\newtheorem{lemma}{Lemma}
\newtheorem{proposition}{Proposition}
\newtheorem{corollary}{Corollary}
\theoremstyle{definition}
\newtheorem{assumption}{Assumption}

\theoremstyle{remark}
\newtheorem{remark}{Remark}

\newcommand{\dist}{\operatorname{dist}}
\newcommand{\Qhat}{\widehat{Q}}
\newcommand{\indic}{\mathbf{1}}

\begin{document}

\title{\textcolor{blue}{REWEAVE}: Online Learning for Workload-Adaptive HNSW Indexes}

\author{Minghao Li}
\affiliation{\institution{University of Toronto}\city{Toronto}\country{Canada}}
\author{Nick Koudas}
\affiliation{\institution{University of Toronto}\city{Toronto}\country{Canada}}
\author{Raghav Mittal}
\affiliation{\institution{The University of Texas At Arlington}\city{Arlington}\country{USA}}

\begin{abstract}
Vector databases rely on Hierarchical Navigable Small World (HNSW)
indexes for similarity search, but a graph built from data alone can be
poorly matched to a skewed or changing query workload. We present
\textcolor{blue}{\textsc{REWEAVE} (Reward-Guided Rewiring of Vector Graphs)}, an online index-maintenance method that learns edge
utility from query feedback and periodically rewires the bottom layer
within the original per-node degree cap. The method supports sparse
exact-neighbor feedback combined with self-labels, and standard HNSW
serving without learning overhead once adaptation is disabled.

Under explicit assumptions %
we prove value-estimation and
graph-stabilization guarantees. %
Under stationary skew, \textcolor{blue}{\textsc{REWEAVE}} improves held-out recall by up to
$23.3$ percentage points over HNSW. In comparisons with query-aware
baselines on up to $100$M vectors, it achieves up to $3.3\times$ NGFix's query throughput at
recall@$100$ of $0.80$. On Amazon-ESCI with query drift, it increases stream
recall@$100$ from $0.745$ to $0.979$ with $34\%$ fewer search distance
computations than frozen HNSW. These results illustrate the potential to
improve retrieval quality and reduce search effort within an existing
graph-degree budget, while incurring additional work during adaptation.

\end{abstract}

\maketitle

\vldbtopmatter

\section{Introduction}
\label{sec:intro}

Approximate nearest-neighbor (ANN) search underpins vector databases~\cite{pan2024survey}, retrieval-augmented generation~\cite{lewis2020rag}, recommendation~\cite{feng2022reinforcementrouting}, and similarity joins~\cite{lu2012knnjoin}. Graph indices lead the recall-latency frontier; HNSW~\cite{malkov2020hnsw} routes queries through progressively denser proximity-graph layers. These indices are built from the corpus alone, although the query workload determines which edges are useful.

Workload-adaptive~\cite{idreos2007databasecracking} and workload-optimized
database systems~\cite{pavlo2017selfdriving, olma2017slalom} adapt physical
design to the queries they serve. Learned indices such as ALEX and the
PGM-index similarly demonstrate the value of updatable, learning-augmented
structures~\cite{ding2020alex, ferragina2020pgm}. For graph ANN, this
motivates using query feedback to decide which edges occupy the index's
degree budget.

Query workloads can differ substantially from the indexed data. Under
\emph{stationary skew}, a fixed, nonuniform query distribution concentrates
on a small region of the data manifold~\cite{jaiswal2022ooddiskann, chen2024roargraph}.
A retrieval-augmented generation system, for example, may index an entire
knowledge base but serve many queries about a few topics. Under
\emph{drift}, interests move the query mass over time, and previously useful
edges can lose value. In both regimes, a corpus-built graph may spend its
degree budget on rarely traversed edges while missing useful links,
reducing recall or increasing search effort.

\iftechreport
\begin{figure}[t]
    \centering
    \includegraphics[width=0.45\linewidth]{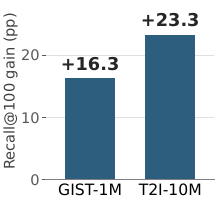}
    \captionsetup{justification=raggedright,singlelinecheck=false}
    \caption{\textcolor{blue}{Held-out Recall@100 gain over frozen HNSW, same degree cap (\S\ref{sec:eval-combinatorial}).}}
    \label{fig:recall-teaser}
    \Description{Recall gains over frozen HNSW: 16.3 percentage points on GIST-1M and 23.3 on T2I-10M.}
\end{figure}
\else
\begin{wrapfigure}{r}{0.40\linewidth}
    \centering
    \includegraphics[width=\linewidth]{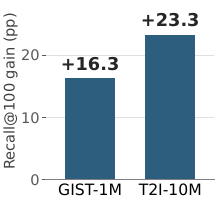}
    \captionsetup{justification=raggedright,singlelinecheck=false}
    \caption{\textcolor{blue}{Held-out Recall@100 gain over frozen HNSW, same degree cap (\S\ref{sec:eval-combinatorial}).}}
    \label{fig:recall-teaser}
    \Description{Recall gains over frozen HNSW: 16.3 percentage points on GIST-1M and 23.3 on T2I-10M.}
\end{wrapfigure}
\fi
Adding edges incurs storage, distance-evaluation, and cache costs. We
therefore keep the per-node out-degree cap $M$ fixed and learn which
destinations serve the workload within that budget. Existing approaches
adapt entry points without changing edges~\cite{oguri2024entrypoint},
build query-aware graphs offline~\cite{chen2024roargraph,ruan2025gate,jaiswal2022ooddiskann},
choose among existing edges~\cite{feng2022reinforcementrouting}, or
incrementally repair or construct graphs from query neighborhoods~\cite{ngfix2026,coral}.
None of these methods mutate the graph under a fixed budget while providing
a regret guarantee against an explicit in-budget benchmark.

Our analysis separates learning to navigate a given graph from discovering
and selecting its edges. This distinction connects adaptation cost to the
supply of useful candidates, while our evaluation measures recall and
serving cost under stationary skew and drift.

We propose \textcolor{blue}{\textsc{REWEAVE} (Reward-Guided Rewiring of Vector Graphs)}, which learns to reallocate a fixed per-node
degree budget to the query workload while serving queries. \textcolor{blue}{\textsc{REWEAVE}} treats navigation as an online learning problem: each outgoing edge is a possible search move, whose reward measures its usefulness on the query. We call this adaptive procedure the
\emph{learner}: it chooses search moves, updates edge scores from query
feedback, and selects edges at rewiring batches. It executes a five-step
loop (Figure~\ref{fig:fig1}): navigate, assign reward, update values, collect candidates, and rewire.

\textcolor{blue}{Figure~\ref{fig:recall-teaser} shows the recall gains workload-driven rewiring can achieve within the original degree budget. They motivate}
our analysis of when feedback reliably identifies useful edges and how
exploration and graph adaptation contribute to learning cost.

Mixed exact/self feedback reduces labeling cost and can recover the
exact-reference graph under per-edge bias and margin conditions
(\S\ref{sec:tworate}). Once learning \textcolor{blue}{stops}, the adapted index serves
with standard HNSW and no learning overhead.
Our core contributions \textcolor{blue}{are:}
\begin{itemize}
    \item A formulation and five-step algorithm for workload-adaptive HNSW, combining edge-reward learning, bounded candidate pools, and degree-preserving rewiring.
    \item Conditional estimation and graph-stabilization guarantees, plus a regret decomposition against a value-greedy graph-and-policy benchmark with logarithmic UCB routing cost and extensions to changing workloads.
    \item An evaluation of navigation, reward, and candidate-generation choices, together with mixed exact/self feedback and standard HNSW serving after adaptation.
\end{itemize}
Code artifacts will be released upon paper acceptance.

\section{Background and Problem Formulation}
\label{sec:formulation}

\subsection{HNSW and the greedy walk}
HNSW~\cite{malkov2020hnsw} stores the corpus
$X=\{x_1,\ldots,x_n\}\subset\mathbb R^d$ at layer~0 and geometrically
sparser subsets at higher layers. With construction parameter $M_{\rm HNSW}$,
the usual degree caps are $2M_{\rm HNSW}$ at layer~0 and $M_{\rm HNSW}$ above.
\textcolor{blue}{Throughout,} $M=2M_{\rm HNSW}$ denotes the \emph{layer-0 cap}; experiments report
$M_{\rm HNSW}$ explicitly. Let $N_{\rm out}(u)$ be the set of outgoing
neighbors of node $u$. A distance-ranked single walk starts at $s$ and
repeatedly visits the closest outgoing neighbor that is strictly closer to
the query than the current node, stopping when no such neighbor exists. Upper-layer descent supplies the layer-0
entry. Standard HNSW serving uses its extended search list of size $ef_{\rm search}$;
our analyzed training walk accepts at most $H$ edges per query.
All distance and score ties use a fixed deterministic rule.

\subsection{Online setting and feedback budgets}
\label{sec:feedback}
Queries $q_1,\ldots,q_T$ arrive sequentially. Our stationary analysis assumes
independent draws from an arbitrary distribution $\mathcal D$; under drift,
draws are independent with phase-specific distributions. These distributions
need not be uniform: they may be skewed and have correlated vector
coordinates. Independence concerns successive queries.
Let $N_k(q)$ be the true top-$k$ set and $A(q)$ the returned set. Recall is
$\mathrm{Recall}_k(q)=|A(q)\cap N_k(q)|/k$. Serving a query means searching
the current graph and returning $A(q)$ to the application. After returning
the answer, the learner uses a feedback neighbor set $L(q)$ to update edge
scores and propose edges for later queries. With exact feedback,
$L(q)=N_k(q)$; with self-feedback, $L(q)=A(q)$, reusing the search's own
answer as its label set (\S\ref{sec:cheapfeedback}). For a directed edge $(u,v)$,
$u$ is the source node and $v$ the destination node. Exact-feedback availability is
distinct from the number of edge rewards recorded:
\begin{itemize}[leftmargin=1.2em,topsep=2pt,itemsep=1pt]
\item \textbf{Out-list full information} scores every current out-edge of each
visited node, costing $O(M)$ membership tests per visit. The stronger
\emph{universe-scoring} regime scores all candidates in a fixed comparison
universe, including nonedges; it supports the full-information analysis
identified in \S\ref{sec:analysis}.
\item \textbf{Bandit feedback} records rewards only on accepted edges:
edges selected by navigation to move to a neighbor closer to the query
than the current node. Direct endpoint or recall credit costs $O(1)$ per
accepted edge. A buffered nonedge receives no
bandit observation merely by being proposed.
\end{itemize}
Exact neighbor search can use certification methods~\cite{li2026hnswcertification,li2025app2exa}.
The analysis of direct endpoint and recall rewards uses exact labels; \S\ref{sec:tworate}
handles mixed feedback under reward-specific conditions.

\subsection{State, action, reward, and edge value}
\label{sec:value}
Each query is one search episode. Before query $t$, the learner's graph
$G_t$ and edge scores reflect preceding queries and remain fixed during
the search. A \emph{navigation policy}
$\pi$ specifies how the search chooses its next outgoing edge, using the
query and information available at the current node; it may randomize
these choices. Write $\pi_t$ for the learner's policy on query $t$.
For a graph $G$ and policy $\pi$, let $\mathcal V_{G,\pi}(q)$ denote the
visited nodes and $P_{G,\pi}(q)$ the ordered sequence of accepted directed
edges. We omit the subscripts when the graph and policy are clear.
No destination node is accepted twice. For a single path,
$P(q)=((u_0,u_1),\ldots,(u_{m-1},u_m))$, $m\le H$.
The endpoint reward is
\begin{equation}
 r_{\rm end}(u,v,q)=\indic[v\in N_k(q)].
 \label{eq:binreward}
\end{equation}
For a fixed graph and policy $(G,\pi)$, its visit-conditional value is
\begin{equation}
 Q_{\rm vis}^{G,\pi}(u,v)
 =\Pr[v\in N_k(q)\mid u\in\mathcal V_{G,\pi}(q)].
 \label{eq:qstar}
\end{equation}
\textcolor{blue}{It measures how often the destination $v$ is a true neighbor among queries visiting $u$.} During learning, we estimate an edge's
reward value from the queries on which the feedback rule evaluates it
(\S\ref{sec:reward}).

\subsection{The bounded candidate pool}
\label{sec:pool}
At time $t$, node $u$ stores its current out-neighbor set
$N_{{\rm out},t}(u)$ and two proposal buffers. The neighbor buffer
$B_{nn,t}(u)$ collects proposed destinations for edges from $u$, using the
feedback neighbor sets $L(q)$ of queries visiting $u$. When exact feedback
provides the true top-$k$ set, $L(q)=N_k(q)$, the missed-neighbor buffer
$B_{miss,t}(u)$ retains proposals from true neighbors that the search did
not examine. Section~\ref{sec:candidates} specifies their update rules.
These sets form the candidate pool
$\mathcal P_t(u)=N_{{\rm out},t}(u)\cup B_{nn,t}(u)\cup B_{miss,t}(u)$
of size at most $C\ge M$. This bounds storage for the current pool, not the number
of distinct candidates encountered over time. For analysis, define
$\mathcal U(u)$ as a fixed finite set of possible destination nodes for
edges from $u$. It contains every candidate that can enter $u$'s pool
during learning and excludes $u$ itself, so
$\bigcup_t\mathcal P_t(u)\subseteq\mathcal U(u)\subseteq X\setminus\{u\}$.
This \emph{comparison universe} specifies the alternatives against which
we assess rewiring, including candidates not stored at the same time.
Retaining reward statistics for candidates removed from the pool requires
storage beyond the $C$ current entries; \S\ref{sec:analysis} accounts for
this history separately.

\subsection{Benchmark and learning objectives}
\label{sec:objective}\label{sec:three_problems}
We measure learning progress against a reference graph and navigation
policy that use fixed edge-value references directly. They obey the same
degree cap and accept only distance-improving moves. For any graph and
policy $(G,\pi)$, the collected endpoint reward and its expectation are
\[
 \begin{aligned}
 \rho(q;G,\pi)&=\sum_{(u,v)\in P_{G,\pi}(q)}\indic[v\in N_k(q)],\\
 R(G,\pi)&=\mathbb E[\rho(q;G,\pi)].
 \end{aligned}
\]
Here $0\le\rho\le\min\{H,k\}$; $\rho/k$ need not equal returned-set recall,
because the search can examine useful points without accepting their edges.

A \emph{credit rule} $r$ specifies the numerical reward $r(e,q)$ assigned
to an edge $e$ for query $q$. Endpoint credit rewards an edge whose
destination is a true neighbor; recall credit assigns the completed
search's recall to each traversed edge (\S\ref{sec:reward}). The rule
therefore determines which aspect of search usefulness the edge scores
learn to reflect.
For each candidate edge $(u,v)$, let $\theta_r(u,v)$ be a fixed reference
value for its average credit under rule $r$. These values define a stable
ranking for comparison: the benchmark uses them directly, while the
learner ranks edges using estimates built from query feedback.
Section~\ref{sec:converge} specifies how much an edge's conditional mean
reward may vary around its reference as the graph and policy adapt.

Let $G^\dagger_r$ retain the top
$\min\{M,|\mathcal U(u)|\}$ candidates by $\theta_r$ at each node. Its policy
$\pi^\dagger_r$ always chooses the highest-$\theta_r$ improving out-edge.
This benchmark's expected collected endpoint reward per query is
$R^\dagger_r=R(G^\dagger_r,\pi^\dagger_r)$.
The subscript $r$ is suppressed when the reward rule is fixed.
These local value rankings need not maximize collected reward or recall
over all degree-$M$ graphs. Changing
from endpoint to recall credit changes the reference graph and policy,
while $R$ continues to measure collected endpoint reward.
The extension to searches that explore several routes uses its own reference
policy and edge values (\S\ref{sec:beam}).

We study three tasks: estimating the edge reward references
$\theta_r(u,v)$, replacing outgoing edges \textcolor{blue}{while keeping} at most $M$ per
node, and adapting these choices as the query distribution changes.
With $R_t=\rho(q_t;G_t,\pi_t)$, expected regret is the deterministic
quantity
\[
 \mathrm{Reg}(T)=T\,R^\dagger-\sum_{t=1}^T\mathbb E[R_t].
\]
This is the benchmark's expected total endpoint reward over $T$ queries
minus the learner's.

\section{The \texorpdfstring{\NoCaseChange{\textcolor{blue}{REWEAVE}}}{REWEAVE} Algorithm}
\label{sec:algorithm}
\begin{figure}
    \centering
    \includegraphics[width=\linewidth]{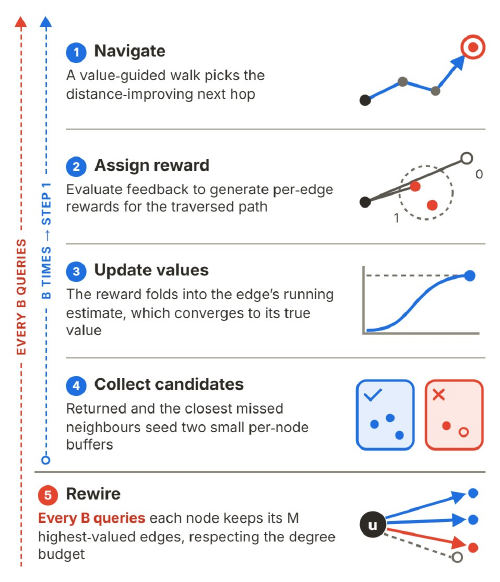}
    \caption{\textcolor{blue}{REWEAVE} as an online loop of five steps.}
    \label{fig:fig1}
\end{figure}

Algorithm~\ref{alg:qlrewire} implements the five steps of Figure~\ref{fig:fig1} for a single path. It starts from
$G_0$ with neighbor count $k$, degree and pool caps $M,C$, rewire interval
$B$, and path cap $H$ (lines~\ref{algline:input}--\ref{algline:init}).
Each edge stored in the current candidate pool receives score $Q_0$ and evaluation count zero;
proposal buffers start empty. An optional beam variant can explore several
routes within a training query (\S\ref{sec:beam}).

For each query, upper-layer descent chooses the entry node $u$
(lines~\ref{algline:query}--\ref{algline:entry}). The set $\mathit{vis}$
records visited nodes, and $P$ records accepted edges in order. A node is
\emph{examined} when the search computes its distance to the query; denote
the set of these nodes by $\mathcal X_{\rm eval}(q_t)$. A node is
\emph{visited} when it is the entry node or the destination of an accepted
edge. Examining a node therefore need not add it to $\mathit{vis}$.
The answer $A(q_t)$ retains the $k$ closest examined nodes, or all examined
nodes if fewer than $k$ are available, using the fixed distance tie rule.
Lines~\ref{algline:rank}--\ref{algline:advance} rank the current node's
out-edges, select the first destination closer to the query, append that
edge to $P$, and continue from its destination. The walk stops when no
out-edge improves distance or $|P|=H$ (lines~\ref{algline:walk},~\ref{algline:stop}).

Line~\ref{algline:answer} serves this answer before feedback or graph
updates. The learner then obtains feedback $L(q_t)$, assigns
rewards under the chosen credit rule, and updates scores and counts
(lines~\ref{algline:labels}--\ref{algline:values}). Bandit feedback evaluates
accepted edges; out-list full information evaluates all current out-edges
at visited nodes. Line~\ref{algline:buffers} then updates the proposal
buffers for visited nodes using the feedback, including missed-neighbor
proposals when exact labels are available. New candidates start at
$(Q_0,0)$; proposing an edge does not count as observing its reward.

Every $B$ queries, lines~\ref{algline:batch}--\ref{algline:rewire} change
the graph. For each node, they unite current out-edges with both proposal
buffers, retain at most $C$ candidates including current outgoing neighbors, and install
the best at most $M$ by score. Distance from the node, then ID, break ties.
Steps 1--4 run for every query; Step 5 runs every $B$ queries. The scores
updated in Step 3 influence navigation on the next query. The proposals
collected in Step 4 influence the graph at the next rewiring batch, whose
out-lists remain in use until another batch. The guide accompanying
Algorithm~\ref{alg:qlrewire} links each step to its pseudocode lines and
detailed treatment.

\begin{algorithm}[t]
\caption{\textcolor{blue}{\textsc{REWEAVE}}, layer 0 (single path).}
\label{alg:qlrewire}
\begin{algorithmic}[1]
\State \textbf{Input:} graph $G_0$; query stream; $k,M,C,B,H$;
navigation rule; reward rule; prior $Q_0$.\label{algline:input}
\State Initialize stored candidate scores to $Q_0$, evaluation counts to $0$,
and proposal buffers to empty.\label{algline:init}
\For{$t=1,2,\ldots$}\label{algline:query}
  \Statex \hspace{\algorithmicindent}\textbf{Step 1: Navigation} (\S\ref{sec:navigate})
  \State $u\gets$ upper-layer descent; $\mathit{vis}\gets\{u\}$; $P\gets[\,]$\label{algline:entry}
  \While{$|P|<H$}\label{algline:walk}
    \State Rank $N_{{\rm out},t}(u)$ by the rule of \S\ref{sec:navigate}\label{algline:rank}
    \State $v\gets$ first ranked neighbor with $\dist(q_t,v)<\dist(q_t,u)$
    \If{no such $v$ exists}\label{algline:stop} \textbf{break} \EndIf
    \State Append $(u,v)$ to $P$; add $v$ to $\mathit{vis}$; $u\gets v$\label{algline:advance}
  \EndWhile\label{algline:walkend}
  \State Serve $A(q_t)\gets$ up to $k$ closest nodes in $\mathcal X_{\rm eval}(q_t)$\label{algline:answer}
  \Statex \hspace{\algorithmicindent}\textbf{Step 2: Reward assignment} (\S\ref{sec:reward})
  \State Obtain $L(q_t)$ after serving the answer\label{algline:labels}
  \State Assign rewards to the pairs permitted by the feedback budget\label{algline:rewards}
  \Statex \hspace{\algorithmicindent}\textbf{Step 3: Value estimation} (\S\ref{sec:estimator})
  \State Update their scores and counts by \eqref{eq:rm} or discounting\label{algline:values}
  \Statex \hspace{\algorithmicindent}\textbf{Step 4: Candidate generation} (\S\ref{sec:candidates})
  \State Update $B_{nn}$ and $B_{miss}$ from $\mathit{vis}$ and the feedback\label{algline:buffers}
  \Statex \hspace{\algorithmicindent}\textbf{Step 5: Rewiring} (\S\ref{sec:rewire-step})
  \If{$t\bmod B=0$}\label{algline:batch}
    \For{each node $u$}
      \State Form $\mathcal P_t(u)=N_{{\rm out},t}(u)\cup B_{nn,t}(u)\cup B_{miss,t}(u)$
      \State Retain at most $C$ pool entries, including current outgoing neighbors
      \State Set $N_{{\rm out},t+1}(u)$ to the best $\min\{M,|\mathcal P_t(u)|\}$
      entries by score, then distance from $u$, then ID\label{algline:rewire}
    \EndFor
  \EndIf\label{algline:batchend}
\EndFor
\end{algorithmic}

\smallskip\noindent\textbf{Guide to the five steps.}\par\smallskip
\noindent{\small
\setlength{\tabcolsep}{3pt}
\begin{tabular}{@{}p{0.40\linewidth}p{0.13\linewidth}p{0.40\linewidth}@{}}
\toprule
Step and section & Lines & Output \\
\midrule
1. Navigation (\S\ref{sec:navigate}) & \ref{algline:entry}--\ref{algline:answer} & Answer $A(q_t)$, accepted edges $P$, visited nodes $\mathit{vis}$ \\
2. Reward assignment (\S\ref{sec:reward}) & \ref{algline:labels}--\ref{algline:rewards} & Per-edge reward observations \\
3. Value estimation (\S\ref{sec:estimator}) & \ref{algline:values} & Updated scores and counts \\
4. Candidate generation (\S\ref{sec:candidates}) & \ref{algline:buffers} & Updated proposal buffers \\
5. Rewiring (\S\ref{sec:rewire-step}) & \ref{algline:batch}--\ref{algline:batchend} & Updated outgoing-edge lists \\
\bottomrule
\end{tabular}}\par\smallskip
\end{algorithm}

Universe-scoring extends reward updates to all $\mathcal U(u)$ at each
visited node, at $O(|\mathcal U(u)|)$ cost. The retained-evidence guarantees
in \S\ref{sec:analysis} require keeping reward statistics and counts across
graph and pool evictions. Under bandit feedback, testing buffered nonedges
requires a separate trial mechanism.

\subsection{Multi-layer integration}
\label{sec:multilayer}
Only layer~0 is adapted. Upper layers retain their distance-based routing
and seed the training walk or beam. After training, frozen serving uses
standard HNSW search on the resulting graph \textcolor{blue}{without} value-update overhead.
The analysis compares training searches with the reference graph and
policy of \S\ref{sec:objective}; the experiments measure the
resulting index's serving performance.

\section{Step 1: Navigation}
\label{sec:navigate}
\label{sec:nav-rejected}
Step 1 (Algorithm~\ref{alg:qlrewire},
lines~\ref{algline:entry}--\ref{algline:answer}) uses the current graph and
scores to return an answer and the search records used by Steps 2--4.
Navigation favors high-reward edges while testing uncertain alternatives:
an early low estimate can otherwise prevent a useful edge from receiving
more feedback. We use $\epsilon$-greedy or UCB ranking, following the first
distance-improving edge in either order. Strict distance improvement
prevents cycles. The optional beam variant (\S\ref{sec:beam}) explores
several routes at a time.

\subsection{\texorpdfstring{$\epsilon$}{epsilon}-greedy Ranking}
\label{sec:eps-rule}
For query $t$, let $\widehat Q_{t-1}(u,v)$ be the score of edge
$(u,v)$ computed from earlier feedback (updates are in \S\ref{sec:estimator}).
\textcolor{blue}{Let $\epsilon_t\in[0,1]$ be the exploration probability.} At each visited
node, with probability $1-\epsilon_t$, order its out-edges by decreasing
score, breaking ties by distance to the query and then ID. With probability
$\epsilon_t$, use a uniformly random order instead. \textcolor{blue}{Either way,} follow
the first edge whose destination is closer to the query than the current
node; stop if none exists.

We use $\epsilon_t=\min\{1,c_\epsilon/\sqrt t\}$, where $c_\epsilon>0$
controls the amount of exploration. Random ordering becomes less frequent
as more queries arrive. Each visited node makes a fresh independent choice
using the same $\epsilon_t$ for that query. Random ordering treats all
improving edges equally, giving each an exploration-selection probability
of at least $\epsilon_t/M$ at that visit. Repeated sampling also requires
queries to visit the node while the edge is present and improving.

\subsection{Optimistic (UCB) Ranking}
\label{sec:ucb-rule}
For a current out-edge $e=(u,v)$, let $N_{t-1}(e)$ count its reward
observations before query $t$, and $\widehat Q_{t-1}(e)$ be their mean.
For direct endpoint or recall rewards in $[0,1]$, upper confidence bound (UCB) navigation ranks
edges by
\[
 U_t(e)=\widehat Q_{t-1}(e)+\sqrt{2\log(t\vee2)/N_{t-1}(e)},
\]
where $t\vee2=\max\{t,2\}$ and $\log$ is the natural logarithm.
The empirical mean favors edges that have earned high rewards. The
square-root term is an \emph{exploration bonus} for sampling uncertainty:
fewer observations provide less evidence and produce a larger bonus.
This lets an uncertain edge compete with one whose observed mean is
higher.

At fixed $t$, four times as many observations halve the bonus. As evidence
accumulates, the ranking therefore relies less on the uncertainty
allowance. Conversely, if an edge's count stays unchanged, the slowly
growing $\log(t\vee2)$ term raises its bonus as queries arrive. The bonus
affects which edge is tried; score updates still use the recorded rewards.

Set $U_t(e)=+\infty$ when $N_{t-1}(e)=0$, giving unobserved current edges
priority if they improve distance. Navigation follows the first improving
edge in index order. This rule ranks only current out-edges; a buffered
nonedge must first be installed before navigation can select it.
Under suitable assumptions detailed in
\S\ref{sec:analysis}, we show that the routing contribution to regret is
$O(\log T)$.

\subsection{Beam Navigation (Width \texorpdfstring{$W$}{W})}
\label{sec:beam}
The single-path walk commits to one next node at each step. Beam navigation
is an optional training variant that keeps several search routes active,
allowing a query to explore different parts of the graph and collect
feedback on more edges. The \emph{frontier} is the set of visited nodes whose
out-edges are ranked at the current step. The beam width $W$ is the
maximum number of nodes in this set.

In the experimental variant, each frontier node proposes its top-$b$
distance-improving neighbors under the chosen ranking rule. Here $b$ is the
\emph{branch factor}, which limits proposals from one node. We pool and
rank these proposals, then accept up to $W$ \textcolor{blue}{destinations} to form the
next frontier. Each destination node is accepted at most once per query:
if several frontier nodes propose the same neighbor, only one proposal is
accepted, avoiding repeated work and duplicate credit. For example, with
$W=4$ and $b=2$, a full frontier proposes at most eight edges, from which
at most four distinct \textcolor{blue}{destinations} form the next frontier.

\textcolor{blue}{A wider beam gathers feedback along more routes at the cost of extra distance evaluations. \S\ref{sec:eval-isolated} compares UCB with $W=1$ and $W=64$ to measure the recall effect.} This choice affects
training navigation; frozen serving uses standard HNSW on the learned graph.

The beam analysis uses all improving out-edges to unvisited destinations,
applying the per-source cap $b$ during acceptance instead of truncating
proposals before pooling. Its guarantees concern this variant, with at most
$H$ total acceptances per query; \S\ref{sec:analysis} identifies the scope
of the extended analysis.

\section{Steps 2 and 3: Reward and Value Estimation}
\label{sec:reward}
\label{sec:reward-rejected}
Step 2 (Algorithm~\ref{alg:qlrewire},
lines~\ref{algline:labels}--\ref{algline:rewards}) combines the completed
search with its feedback label set $L(q)$. This \emph{credit assignment}
records a numerical reward for each edge receiving feedback to reflect
its usefulness on that query. The chosen
reward rule $r$ specifies the value $r(e,q)$ assigned to edge $e=(u,v)$ for
query $q$. Sections~\ref{sec:endpoint} and~\ref{sec:recallcredit} define
direct endpoint and recall credit; \S\ref{sec:gammavariants} describes ways to
credit earlier edges for later search steps. These are choices within
Step 2. Step 3, \emph{value estimation} (line~\ref{algline:values}), then
combines recorded rewards into the edge scores used by navigation and
rewiring; \S\ref{sec:estimator} gives its update rules.

Let $G$ be the layer-0 graph and $\pi$ the navigation policy, which specifies
how the search chooses outgoing edges. The feedback rule determines which
edges receive reward observations; we call these edges \emph{evaluated}.
Holding $G$, $\pi$, and the feedback rule fixed, the expected recorded
reward of an edge is
\begin{equation}
 Q_r^{G,\pi}(e)=\mathbb E[r(e,q)\mid e\text{ is evaluated}],
 \label{eq:qstar-r}
\end{equation}
whenever that edge has positive probability of being evaluated. The
expectation averages over queries and navigation randomness, conditional
on the edge receiving feedback.
For example, bandit feedback averages over queries whose search accepts
the edge. As the graph and policy adapt, the queries evaluating an edge
can change, so this average may change even under a fixed query
distribution. For our guarantees, \S\ref{sec:converge} states the assumption
limiting this variation around the fixed reference $\theta_r(e)$ introduced
in \S\ref{sec:objective}.

\subsection{The Binary Endpoint Reward}
\label{sec:endpoint}
The endpoint reward in \eqref{eq:binreward} is $1$ when an edge's
destination $v$ belongs to the true top-$k$ set $N_k(q)$, and $0$ otherwise.
It measures whether that individual edge leads directly to a true neighbor.

Under \emph{bandit feedback} (\S\ref{sec:feedback}), only traversed edges
in $P(q)$ receive observations. Both $0$ and $1$ increment the evaluation
count and enter the score update; an untraversed edge receives no update.
The expected reward in \eqref{eq:qstar-r} therefore measures how often $v$
is a true neighbor among queries whose search traverses $(u,v)$.

With \emph{out-list full information}, we score every current outgoing
edge of each visited node, including edges the search did not follow.
For fixed $G$ and $\pi$, the average for $(u,v)$ therefore includes all
queries that visit $u$, giving $Q_{\rm vis}^{G,\pi}(u,v)$ in
\eqref{eq:qstar}. These queries can differ from those that traverse the
edge, so the two feedback rules need not produce the same expected reward.

Endpoint reward gives positive credit only for reaching a true neighbor.
\textcolor{blue}{For example, a path whose three destinations lie outside, inside, and inside $N_k(q)$ receives rewards $0,1,1$. The first edge receives zero even if it is needed to reach the two useful nodes after it.} This
limitation motivates the recall reward described next.

\subsection{The Recall Reward}
\label{sec:recallcredit}
The recall reward gives an edge credit for participating in a search that
returns useful results. After query $q$ completes, compute
$\mathrm{Recall}_k(q)$, the fraction of its true top-$k$ neighbors present
in the returned result. Assign this same value to every traversed edge:
$r_{\rm rec}(e,q)=\mathrm{Recall}_k(q)$ for all $e\in P(q)$.

For example, if the search returns $80\%$ of the true neighbors, every
traversed edge receives reward $0.8$. This includes an intermediate edge
whose destination lies outside $N_k(q)$ and would therefore receive zero
endpoint reward. Recall credit can consequently reward earlier steps
along a successful route. Because all traversed edges receive the same
value, it does not distinguish their individual contributions to the result.

Only traversed edges receive observations. \textcolor{blue}{For fixed $G$ and $\pi$, \eqref{eq:qstar-r} therefore measures average recall over searches traversing $e$.} The estimation guarantees require the conditional-mean,
retained-evidence, and coverage conditions of \S\ref{sec:converge} for these
recall rewards.

\subsection{Assigning Credit for Later Search Steps}
\label{sec:gammavariants}
Within Step 2 (Algorithm~\ref{alg:qlrewire}, line~\ref{algline:rewards}),
two credit-assignment variants give earlier edges credit for useful results
reached later in the search. Both start from the endpoint or recall reward
$r(e,q)$ and change the observation used to update the edge score.

Let $\gamma\in[0,1)$ be the \emph{discount factor for later search steps}.
A reward one step later has weight $\gamma$, two steps later has weight
$\gamma^2$, and so on. Setting $\gamma=0$ uses only the current edge's
base reward; larger values give more weight to later rewards. This
discount applies within a query; \S\ref{sec:estimator} separately
describes how old evidence is downweighted across queries.
We consider Monte Carlo (MC) and temporal-difference (TD) credit, denoted
MC($\gamma$) and TD($\gamma$).

\emph{Monte Carlo (MC) credit} uses the rewards actually recorded during
the completed search. For a path of $m$ accepted edges
$e_i=(u_i,u_{i+1})$, $i=0,\ldots,m-1$, assign edge $e_i$ the discounted
return
\[
 Z_i^{(\gamma)}=\sum_{j=i}^{m-1}\gamma^{j-i}r(e_j,q).
\]
This return sums its own reward and the weighted rewards of all later
edges. For the endpoint rewards $0,1,1$ in \S\ref{sec:endpoint}, choosing
$\gamma=0.8$ gives the first edge credit $0+0.8+0.8^2=1.44$, although its
own endpoint reward is zero. We use this return as the observation
$Y_t(e_i)$ in the estimator of \S\ref{sec:estimator}. All returns can be
computed in $O(m)$ time by processing the path backward.

For beam navigation, order $P(q)$ by frontier step and then by acceptance
order within that step. MC uses this ordered sequence. Later entries may
come from different branches, so the return includes later accepted-edge
rewards even when those edges are not descendants of the credited edge.

\emph{Temporal-difference (TD) credit} uses the current edge's reward and
an estimate of the reward available after reaching its destination. For
an accepted edge $(u,v)$, its update target is
\[
 r(u,v,q)+\gamma\max_{w\in N_{{\rm out},t}(v)}\widehat Q(v,w).
\]
Here $N_{{\rm out},t}(v)$ is the current set of outgoing neighbors of $v$;
the maximum selects the largest score among those outgoing edges. The
continuation term is zero when the search terminates, and out-lists remain
fixed during the query. Step 3 uses this target as $Y_t(u,v)$ in the
estimator of \S\ref{sec:estimator}. Using learned scores
inside the target is called \emph{bootstrapping}: TD uses an estimated
continuation, whereas MC uses the observed later rewards. Computing the
maximum costs $O(M)$ per update.

These variants can produce values larger than $1$. For base rewards in
$[0,1]$, MC returns are at most $(1-\gamma)^{-1}$; endpoint rewards give
the tighter bound $\min\{k,(1-\gamma)^{-1}\}$ because accepted destination
nodes do not repeat, including in a beam. TD scores remain in
$[0,(1-\gamma)^{-1}]$ when initialized there and updated by convex
averages. For a chosen credit range $[0,R_{\max}]$, keep the prior on that
scale and multiply the UCB bonus in \S\ref{sec:ucb-rule} by $R_{\max}$.
Scaling the bonus specifies an exploration rule; guarantees
also require the conditional-mean assumptions for the chosen credit.
The regret bounds for base rewards in \S\ref{sec:analysis} do not cover
TD's bootstrapped targets. The estimation result in
\S\ref{sec:cheapfeedback} covers bounded MC returns with running means
under the conditional-mean, retained-evidence, and coverage assumptions
specified there.

\subsection{Running Means and Time-Indexed Discounting}
\label{sec:estimator}
Step 3 (Algorithm~\ref{alg:qlrewire}, line~\ref{algline:values}) takes
the reward observations assigned in Step 2 and updates each evaluated
edge's score and count using running means or discounting. These scores
guide the next query and rewiring batch.
The input $Y_t(e)$ is $r(e,q_t)$ for direct endpoint or recall credit, the
return $Z_i^{(\gamma)}$ for MC, or the bootstrapped target for TD
(\S\ref{sec:gammavariants}).
Let $I_t(e)$ indicate whether $e$ is evaluated on query $t$, so
$N_t(e)=\sum_{s\le t}I_s(e)$ for retained evidence. If $e$ is evaluated
with reward $Y_t(e)$, set
\begin{equation}
 \widehat Q_t(e)=\widehat Q_{t-1}(e)
 +\frac{Y_t(e)-\widehat Q_{t-1}(e)}{N_{t-1}(e)+1},
 \qquad N_t(e)=N_{t-1}(e)+1.
 \label{eq:rm}
\end{equation}
Unobserved entries are unchanged. Thus the first observation replaces $Q_0$,
and subsequent scores are exact running means. The update weight depends
on that edge's observation count, so well-sampled edges change less after
one new reward. Query $t$ uses scores from preceding queries. An endpoint prior is
$Q_0=k/n$; it provides initialization, not evidence of a candidate's quality.

Under \emph{query-workload drift}, the query distribution changes over
time, so older observations can become less representative. A running mean
weights all rewards equally, so a large history can slow adaptation.
\textcolor{blue}{Time-indexed discounting favors recent queries.}

Choose a forgetting factor $0<\lambda<1$: an observation received $j$
queries ago has weight $\lambda^j$. A value close to $1$ retains older
evidence longer; a smaller value emphasizes more recent feedback. For each
edge $e$, maintain a discounted reward sum $S_t^\lambda(e)$ and a
discounted observation count $N_t^\lambda(e)$. Initialize both to zero and,
after each query $t$, update
\[
 \begin{aligned}
 S_t^\lambda(e)&=\lambda S_{t-1}^\lambda(e)+I_t(e)Y_t(e),\\
 N_t^\lambda(e)&=\lambda N_{t-1}^\lambda(e)+I_t(e),\\
 \widehat Q_t^\lambda(e)&=S_t^\lambda(e)/N_t^\lambda(e).
 \end{aligned}
\]
The ratio is a weighted average of recorded rewards; at zero count,
use $Q_0$. Both statistics decay with every arriving query, including
\textcolor{blue}{those} that do not evaluate $e$. An implementation can apply this decay
when the entry is next accessed by multiplying both statistics by
$\lambda$ raised to the number of elapsed queries. If no new observation
arrives, numerator and denominator shrink by the same factor, so the score
stays fixed. \textcolor{blue}{The declining count nevertheless reduces the evidence supporting that score.} The forgetting timescale is
$O(1/(1-\lambda))$ queries, while storage remains constant per retained edge.

Discounted UCB combines these estimates with an exploration bonus based
on the discounted observation count. When an edge goes unobserved, its
count decays and its bonus increases, encouraging the search to revisit
edges whose values may have changed. Its analysis under changing workloads
is among the extensions identified in \S\ref{sec:analysis}.

\section{Step 4: Candidate Generation}
\label{sec:candidates}
\label{sec:cand-rejected}
After serving each query, Step 4 (Algorithm~\ref{alg:qlrewire},
line~\ref{algline:buffers}) updates the two proposal buffers of
\S\ref{sec:pool}, using visited nodes from Step 1 and feedback $L(q)$ from
Step 2. A proposed destination $v$ at node $u$ represents the possible edge
$(u,v)$. At the next batch, proposals join current outgoing neighbors in
$\mathcal P_t(u)$ and compete by score for selection in Step 5. Capacity $C$ counts distinct destinations in
this union, so a node appearing in both buffers occupies one pool entry.
Keeping the current outgoing neighbors leaves at most
$C-|N_{{\rm out},t}(u)|$ entries for distinct new destinations. The
allocation between proposal buffers must respect this shared budget;
the graph-stabilization guarantee separately requires the candidate
availability stated in \S\ref{sec:converge}.

\subsection{The Neighbor Buffer}
The neighbor buffer $B_{nn}(u)$ stores recent feedback neighbors of
queries that visit $u$. We use destinations in $L(q)$, excluding $u$
itself, to insert new candidates and refresh existing entries. The
buffer follows a least-recently-used (LRU) rule: when space is needed,
it replaces the entry least recently inserted or refreshed.

Exact feedback supplies candidates from all true top-$k$ neighbors,
including missed ones; self-feedback supplies the returned nodes, as
defined by $L(q)$ in \S\ref{sec:feedback}.

\subsection{The Missed-Neighbor Buffer}
\label{sec:missbuffer}
Exact feedback also identifies useful destinations that the search missed.
Among the true top-$k$ neighbors outside $\mathcal X_{\rm eval}(q)$, the
examined-node set defined in \S\ref{sec:algorithm}, we
select the one closest to $q$, when such a neighbor exists. We insert this
destination into the missed-neighbor buffer $B_{miss}(u)$ at every visited
node $u$, proposing a direct edge from that node to the missed neighbor.
This buffer tracks proposals generated specifically by missed results
within the bounded candidate storage. A destination may occur in both
buffers: $B_{nn}$ records it as a recent feedback neighbor, while
$B_{miss}$ records its role in a missed search result. Rewiring considers
the destination once, using its shared edge score.

\section{Step 5: Budget-Preserving Rewiring}
\label{sec:rewire-step}
Step 5 (Algorithm~\ref{alg:qlrewire},
lines~\ref{algline:batch}--\ref{algline:batchend}) \textcolor{blue}{uses the scores from Step 3 and proposals from Step 4} to rebuild outgoing-edge
lists every $B$ queries. For node $u$, combine its current outgoing neighbors and
the two proposal buffers, retaining at most $C$ distinct candidates,
including the current \textcolor{blue}{ones.} The shared capacity $C$ is
enforced as in \S\ref{sec:candidates}; at most $M\le C$ candidates become
outgoing edges.

Rank the candidates in $\mathcal P_t(u)$ by their edge scores
$\widehat Q(u,v)$, breaking ties first by distance from $u$ and then by
node ID. Install edges to the best $\min\{M,|\mathcal P_t(u)|\}$
destinations, using scores from \S\ref{sec:estimator}. The resulting
outgoing-edge list is used until the next batch.
Sorting costs $O(C\log C)$ per node per rewiring batch, and selecting at
most $M$ destinations preserves the degree cap.

Optional reverse proposals
let selecting $(u,v)$ propose $(v,u)$ at node $v$. The recipient ranks it
by its own score $\widehat Q(v,u)$, subject to the same pool and degree caps
and top-$M$ selection. The stabilization conditions in \S\ref{sec:converge} apply with or
without these proposals when the fixed comparison universes include every
possible directed proposal. They require retained evidence and coverage
for that universe; reverse proposals alone do not establish these conditions.
Inserting reciprocal edges after top-$M$ selection would bypass the rule
used by the guarantee.

\section{Theoretical Analysis}
\label{sec:analysis}
The analysis explains when feedback can identify a stable graph and how
quickly navigation learns to use it. We focus on UCB, whose exploration
adapts to the evidence for each edge. The comparison uses the value-greedy
benchmark of \S\ref{sec:objective}, a fixed corpus, and the finite universes
of \S\ref{sec:pool}. Let $V_\circ$ contain every node where the learner or
reference policy can choose an edge, and let
$K=\sum_{u\in V_\circ}|\mathcal U(u)|$ count all compared pairs, including
rarely visited nodes. Retaining their reward statistics requires $O(K)$
storage; $H$ caps total accepted edges per query.

\iftechreport Appendix~\ref{app:proofs8}\else The full version~\cite{qlearntechreport}\fi{} provides proofs and finite-sample
bounds for \S\ref{sec:analysis}--\S\ref{sec:tworate}, analyses of
universe-scoring, $\epsilon$-greedy, discounted UCB under workload changes,
and the full-proposal beam.%

\subsection{When Learned Values Support Reliable Decisions}
\label{sec:converge}
The three assumptions control reward variation, separation between
competing choices, and the supply of evidence and candidates for rewiring.

\begin{assumption}[Bounded conditional means]
\label{ass:means}
Queries are independent from a fixed $\mathcal D$. Write $\mathcal F_{t-1}$
for the complete preceding history. Each evaluated pair $e$ records one
reward $Y_t(e)\in[0,1]$ per query; $I_t(e)$ indicates that evaluation.
There are fixed references $\theta_r(e)\in[0,1]$
and $\kappa\ge0$ such that
\[
 |\mathbb E[Y_t(e)\mid\mathcal F_{t-1},I_t(e)=1]-\theta_r(e)|\le\kappa
\]
whenever evaluation has positive conditional probability. For
out-list full information, evaluation requires that the edge is present
and its source is visited; universe-scoring evaluates every compared edge
at a visited source. Bandit feedback evaluates accepted edges.
\end{assumption}
This gives each edge a fixed reference for learning while allowing its
observed reward mean to change as navigation adapts. The tolerance
$\kappa$ bounds variation in the conditional mean, rather than sampling
noise that vanishes with more observations. The
assumption is required separately for the chosen endpoint or recall reward.

\begin{assumption}[Separate selection margins]
\label{ass:margins}
Let $0<\Delta_{\rm set},\Delta_{\rm nav}\le1$. At every node with
$|\mathcal U(u)|>M$, the $M$-th largest reference value exceeds the
$(M+1)$-th largest by at least $\Delta_{\rm set}>2\kappa$.
For single-path navigation on $G^\dagger$, every nonempty set of improving
out-edges that can occur has a unique highest-reference edge, separated
from every other by at least $\Delta_{\rm nav}>2\kappa$.
Put $g_{\rm set}=\Delta_{\rm set}-2\kappa$ and
$g_{\rm nav}=\Delta_{\rm nav}-2\kappa$.
\end{assumption}
The set margin protects the choice of outgoing edges; the navigation
margin protects the choice among edges that improve distance for the
current query. Each of two compared means can move toward the other by
$\kappa$, reducing their separation by at most $2\kappa$. The positive
remaining gap ($g_{\rm set}$ or $g_{\rm nav}$) therefore preserves their
order once each estimate's sampling error is below half that gap.
For example, a reference gap of $0.20$
with $\kappa=0.03$ leaves a gap of $0.14$ for sampling to distinguish.

\begin{assumption}[Coverage for graph stabilization]
\label{ass:coverage}
Every pair in the comparison universes is evaluated infinitely often,
its reward statistics and observation count are retained even when it is
removed from the graph or candidate pool, and every edge of
$G^\dagger$ is present in its source node's pool at every sufficiently late batch.
\end{assumption}
Continued evaluation and retained evidence support accurate estimates;
eventual candidate availability lets rewiring install the reference
edges. For retained running means, Assumption~\ref{ass:means}, the set
margin, and Assumption~\ref{ass:coverage} yield almost-sure stabilization
at $G^\dagger$. The navigation margin is used separately for the routing
bound in \S\ref{sec:regret}.

\subsection{The Main Guarantee and Its Practical Implications}
\label{sec:regret}
To separate candidate discovery from navigation, let
$D_T=\sum_{t\le T}\Pr(G_t\ne G^\dagger)$ count graph mismatch in
expectation, comparing outgoing-edge lists on $V_\circ$. Comparing the
learner's search and the reference search on the same query gives
\begin{equation}
 |\mathrm{Reg}(T)|\le H D_T+H\,\mathbb E[J_T],
 \label{eq:regdecomp}
\end{equation}
where $J_T$ counts queries on which $G_t$ matches $G^\dagger$ on $V_\circ$
but the accepted path differs from the path taken by $\pi^\dagger$.
Each different trajectory can change collected reward by at
most $H$. The absolute bound also permits the learner to outperform this
specified benchmark. The central result controls the second term.

\begin{theorem}[UCB navigation]
\label{thm:ucb}
Under Assumption~\ref{ass:means} and the navigation margin of
Assumption~\ref{ass:margins}, use the UCB index of \S\ref{sec:ucb-rule}
with retained running-mean evidence. Then
\[
 |\mathrm{Reg}(T)|\le HD_T+O\!\left(\frac{K\log(T\vee2)}{g_{\rm nav}^2}\right),
\]
where the constant suppresses the fixed episode cap $H$. The routing
term allows nodes to be visited arbitrarily rarely.
\end{theorem}

The routing term quantifies the cost of learning to use the benchmark
graph. An uncertain edge may initially receive a high index, but accepting
it also supplies an observation. After enough such observations, its
confidence bonus is too small to displace a better improving edge, except
on rare estimation errors. This limits cumulative routing mistakes to
logarithmic growth. Rarely available edges are charged only when a routing choice involving them
occurs. The proof counts the first different choice in each query, so it
does not assume that a local reward gap bounds the loss along the remaining
trajectory.

The graph term \textcolor{blue}{is} the cost of adapting the index: candidate
discovery, evaluation, retention, and rewiring determine when reference
edges are available and selected. We bound $D_T$ independently of $T$
under the conditional-mean and set-margin
conditions of \S\ref{sec:converge}, with retained evidence, all reference
edges available at every batch, and a uniform positive lower bound on
each compared pair's conditional evaluation probability per query.

Combined with Theorem~\ref{thm:ucb}, this gives an $O(\log T/T)$ difference
in average expected training reward for fixed problem parameters. More
generally, $D_T=o(T)$ suffices: the expected fraction of queries using a
different graph tends to zero, so the average reward difference also
vanishes. These results
explain how adaptation and exploration costs amortize over a stationary
workload. \textcolor{blue}{The comparison uses the graph and policy benchmark of \S\ref{sec:objective}; frozen HNSW serving is evaluated empirically in \S\ref{sec:exp}.}

\section{Self-Supervised Feedback with Sparse Exact Labels}
\label{sec:cheapfeedback}\label{sec:tworate}
To reduce the cost of exact-neighbor labels, we combine occasional exact
feedback with \emph{self-labels}: the returned set $A(q)$ serves as the
label set $L(q)$ (\S\ref{sec:feedback}). In this \emph{two-rate feedback}
scheme, each query uses exact feedback with probability $p\in[0,1]$,
independently of the history, query, and search randomness, and self-feedback
otherwise.
Both modes evaluate the same completed-search edges and update the same
running means; exact labels can also supply missed-neighbor proposals.
Expected additional feedback cost is $pc_{\rm ex}+(1-p)c_{\rm self}$ per
query, where $c_{\rm ex}$ and $c_{\rm self}$ include each mode's label-search
and processing costs.

For bounded, nonnegative rewards, suppose each mode's mean, conditional
on prior history and evaluation of edge $e$, stays within $\kappa_j$ of a
fixed reference $\theta_j(e)$, for $j\in\{{\rm ex},{\rm self}\}$, as in
\S\ref{sec:converge}. Here $\theta_{\rm ex}$ and $\theta_{\rm self}$ play the role of
$\theta_r$ for the two feedback modes; $\kappa_j$ is the corresponding
conditional-mean tolerance. The shared empirical score $\widehat Q$
estimates the mixed reference $\theta_p$ defined below. With reward sums and counts retained across rewiring
and eviction, every infinitely evaluated edge's running mean has
almost-sure asymptotic error
at most $\kappa_p=p\kappa_{\rm ex}+(1-p)\kappa_{\rm self}$ about
$\theta_p=p\theta_{\rm ex}+(1-p)\theta_{\rm self}$.

\textbf{Recovering the exact-reference graph.}
Let $b_{\rm label}=\sup_e|\theta_{\rm self}(e)-\theta_{\rm ex}(e)|$ measure the largest
per-edge disagreement between feedback references, and let
$\Delta_{{\rm ex},{\rm set}}$ be the minimum gap between selected and
excluded edges under $\theta_{\rm ex}$, over nodes with a selection boundary.
Under the preceding conditions, if every compared edge in the fixed finite
universes of \S\ref{sec:pool} is evaluated infinitely often and the
exact-reference top-$M$ edges are available at every sufficiently late
batch, the graph stabilizes almost surely at that exact-reference graph
whenever
\begin{equation}
 2[\kappa_p+(1-p)b_{\rm label}]<\Delta_{{\rm ex},{\rm set}}.
 \label{eq:tworate-exact-margin}
\end{equation}
Thus sparse exact
feedback can preserve the selected edges when estimation tolerance and
self-label disagreement remain below the separation between competing
edges. For fixed references, increasing $p$ reduces the self-label
contribution $(1-p)b_{\rm label}$ at the cost of more exact feedback.

In the label-efficiency experiment (\S\ref{sec:eval-tworate}), both modes
record self-labeled MC($0.8$) rewards, while exact labels supply candidate
proposals; this isolates the proposal benefit of exact feedback.

\section{Experimental Evaluation}
\label{sec:evaluations}
\label{sec:exp}

\subsection{Experiment Overview}
\label{sec:eval-overview}

We ablate \textcolor{blue}{\textsc{REWEAVE}}'s three design axes (navigation, reward,
candidate generation) in isolation (\S\ref{sec:eval-isolated}) and jointly
(\S\ref{sec:eval-combinatorial}), study workload locality
(\S\ref{sec:eval-locality}), convergence (\iftechreport\S\ref{sec:eval-convergence}\else\textcolor{blue}{full version~\cite{qlearntechreport}}\fi) and
label efficiency (\S\ref{sec:eval-tworate}), compare against
HNSW~\cite{malkov2020hnsw}, NGFix~\cite{ngfix2026},
NAP~\cite{chen2024roargraph}, CORAL~\cite{coral}, and
CleanANN~\cite{cleann2026}
(\S\ref{sec:eval-system}), and test drift on real shopping queries
(\S\ref{sec:eval-drift}). Datasets are in Table~\ref{tab:datasets}; the
baseline comparison also runs on T2I-100M to test scale, and other 1M-scale
benchmarks behave consistently.

\begin{table}[t]\centering
\footnotesize
\setlength{\tabcolsep}{1pt}
\begin{tabular}{@{}p{0.3\columnwidth}p{0.4\columnwidth}rcc@{}}
\toprule
Dataset & Description & \# vectors & Dim. & Metric \\
\midrule
Amazon-ESCI & shopping product embeddings & $1.44$M & $384$ & cosine \\
GIST-1M & deep image features & $1$M & $960$ & $\ell_2$ \\
T2I-10M & Text2Image multimodal embeddings & $10$M & $200$ & cosine \\
T2I-100M & first $100$M of Text2Image-1B & $100$M & $200$ & cosine \\
\bottomrule
\end{tabular}
\caption{\textcolor{blue}{Datasets. Amazon-ESCI is used only for drift (\S\ref{sec:eval-drift}); T2I-100M only for query-aware baselines (\S\ref{sec:eval-roargraph}).}}
\label{tab:datasets}
\end{table}

\noindent\textbf{Workload 1: Static.} A stationary stream concentrated on a
set of query basins, stressing cross-basin routing. We cluster each dataset
with HDBSCAN\footnote{\url{https://github.com/TutteInstitute/fast_hdbscan}}
and take the median-density cluster $\mathcal C$ (density $(\operatorname{median}_{x\in \mathcal C} \dist_k(x;\mathcal C))^{-1}$,
$\dist_k$ the $k$th-NN distance within $\mathcal C$). Each query is
$q = x + \xi$, $x\in \mathcal C$ sampled uniformly,
$\xi \sim \mathcal{N}\big(0,(\sigma\operatorname{diam}(\mathcal C))^2 I/d\big)$ in dimension $d$, with
disjoint seeds for training and test. We average $\sigma \in \{0.10, 0.25, 0.50\}$
over $3$ seeds each (\S\ref{sec:eval-locality} varies $\sigma$ up to $1$) and,
unless noted, use $1000$ training and $100$ test queries with exact ground truth.

\begin{remark}
\label{rm:label}

Unless otherwise stated, experiments with partial exact feedback use
recall reward and $B_{miss}$ on exact-labeled queries, and self-labeled
MC($0.8$) on the remaining queries (\S\ref{sec:cheapfeedback}).
The label-efficiency experiment (\S\ref{sec:eval-tworate}) uses
self-labeled MC($0.8$) throughout and exact neighbors for $B_{miss}$.
In the comparisons with NGFix, NAP\textcolor{blue}{,} CORAL \textcolor{blue}{and CleanANN}
(\S\ref{sec:eval-roargraph}, \S\ref{sec:eval-drift}), \textcolor{blue}{\textcolor{blue}{\textsc{REWEAVE}}, NGFix, NAP and CORAL}
use the IDs returned by the serving ANN search as self-labels\textcolor{blue}{; CleanANN uses no labels}.
\end{remark}

\noindent\textbf{Workload 2: Drift.} Real Amazon ESCI shopping
queries~\cite{reddy2022shopping}, sampled with replacement from the
base-query pools within category clusters, with category switches
(\S\ref{sec:eval-drift}).
We construct workloads because no public vector-search trace carries
timestamps, and the production trace available to us cannot be released.

All code is C++ on hnswlib\footnote{\url{https://github.com/nmslib/hnswlib}},
run on an Intel Xeon Ice Lake 3.5\,GHz with 247\,GB memory; T2I-100M runs
use a cluster node with 480\,GB memory allocated. Defaults: HNSW
$M_{\rm HNSW}{=}32$, $ef_{\rm construction}{=}200$, $ef_{\rm search}{=}100$, $k{=}100$; rewiring
every $B{=}100$ queries; pool cap $C{=}128$ including
incumbents, in all experiments; prior $Q_0=k/n$. UCB with beam width $W$
is UCB($W$) ($W{=}64$ for every beam run), with branch cap $b{=}32$; discounted rewards are MC($\gamma$) or TD($\gamma$).

\noindent\textbf{Protocol and costs.} Each online query is searched and its
answer served first; recall is measured on that served answer, and only then
does the query's feedback update the edge values and, every $B$ queries, the
graph. No query is re-searched. Recall is always scored against exact
brute-force ground truth, whatever label the learner sees. End-to-end cost
charges search, value updates, label generation, rewiring, drift-detector
maintenance and graph synchronization. Distance computations count every
query--vector distance evaluated while serving.

\subsection{Isolated Axes Testing}
\label{sec:eval-isolated}
Each ablation fixes two axes at the reference setting (first row of
Table~\ref{tab:qlearn-design-space}: $\epsilon$-greedy with $c_\epsilon{=}2$,
endpoint reward, $B_{nn}$ only) and varies the third; beams use $W{=}64$.
\begin{footnotesize}
\begin{table}[t]\centering
\begin{tabular}{lll}
\toprule
Navigation \S\ref{sec:navigate} & Reward assignment \S\ref{sec:reward} & Candidate pool \S\ref{sec:candidates} \\
\midrule
$\epsilon$-greedy & endpoint ($\gamma{=}0$) & $B_{nn}$ \\
UCB($1$) & \textbf{Recall reward} & $\boldsymbol{B_{nn}+B_{miss}}$ \\
\textbf{UCB($64$)} & MC($0.8$), MC($0.1$) & \\
 & TD($0.8$), TD($0.1$) & \\
\bottomrule
\end{tabular}
\caption{\textcolor{blue}{\textsc{REWEAVE}} design space. First row: reference setting; highest-recall method in~\S\ref{sec:eval-combinatorial} in bold \textcolor{blue}{for} each column.}
\label{tab:qlearn-design-space}
\end{table}
\end{footnotesize}

\paragraph{Navigation rule.}
\label{sec:eval-nav}

\begin{figure}[t]\centering

\includegraphics[width=0.8\linewidth]{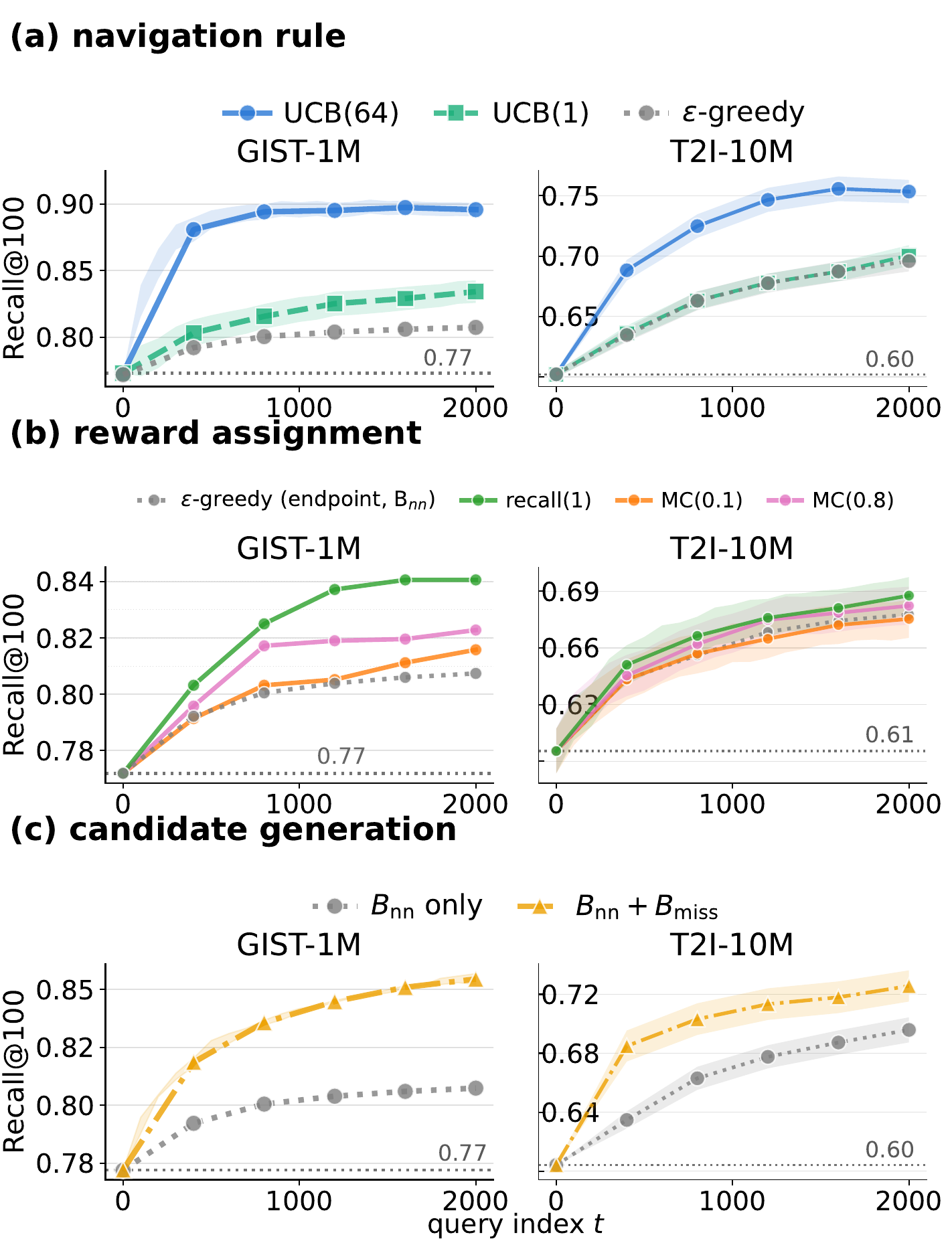}
\caption{Isolated ablations: GIST-1M (left), T2I-10M (right).}
\label{fig:f1_nav_recall}
\label{fig:f3_learning}
\label{fig:f6_candidate}
\end{figure}

In Figure~\ref{fig:f1_nav_recall}(a), UCB(64) beats the second-best UCB(1)
by $6.2$ points on GIST-1M and $5.3$ on T2I-10M at the final checkpoint: a wide
beam exposes more useful routing edges. This comparison changes beam width while keeping UCB ranking, so it measures the value of width; we isolate the learned values below. UCB($1$) stays slightly above
$\epsilon$-greedy, as optimism targets uncertain edges rather than random ones.

\paragraph{Reward assignment.}
\label{sec:eval-reward}

We compare recall reward, discounted binary reward
($\gamma\in\{0.8,0.1\}$, MC or TD backups) and endpoint
reward (\S\ref{sec:endpoint}) with out-list full information.
\begin{wraptable}{L}{6cm}
\centering
\footnotesize
\setlength{\tabcolsep}{4pt}
\begin{tabular}{lcc}
\toprule
variant & GIST-1M & T2I-10M \\
\midrule
TD($0.8$)        & $0.813$ & $0.676$ \\
TD($0.1$)        & $0.810$ & $0.677$ \\
Endpoint (full information) & $0.807$ & $0.647$ \\
\bottomrule
\end{tabular}
\caption{Recall@$100$ at the end of training for reward variants not plotted.}
\label{tab:reward_final}
\end{wraptable}

Every rule lifts recall above the frozen graph (GIST-1M $0.772$, T2I-10M
$0.605$; Figure~\ref{fig:f3_learning}(b), Table~\ref{tab:reward_final}), and
differences between rules are small next to that lift. Only recall beats
the endpoint reward on both datasets ($0.819$ vs.\ $0.816$; $0.688$ vs.\
$0.678$). The larger discount wins under MC ($0.814$/$0.682$ vs.\
$0.810$/$0.675$) and under TD on GIST-1M, plausibly because it propagates
downstream success to upstream routing edges so they are not replaced. Full
information ends $2.9$ points below TD($0.8$) on T2I-10M while changing the
most edges. Unless otherwise stated, we use recall reward with exact
labels and MC($0.8$) with self-labels (Remark~\ref{rm:label}).

\paragraph{Candidate generation.}
\label{sec:eval-candidates}

Adding $B_{miss}$ to $B_{nn}$ (Figure~\ref{fig:f6_candidate}(c)) gains
$4.7$ points on GIST-1M and $3.0$ on T2I-10M at the final checkpoint, by
re-injecting true neighbors the search missed.

\begin{figure}[t]\centering
\includegraphics[width=0.8\linewidth]{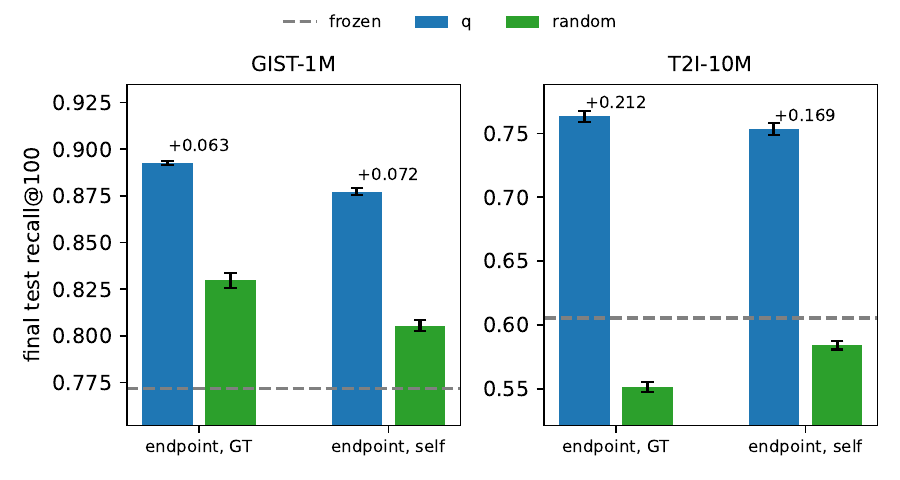}
\caption{\textcolor{blue}{Final test recall@100, learned ($\widehat Q$) vs.\ random $B_{nn}$ ranking (binary endpoint reward, GT/self labels, $3$ seeds). Numbers: learned minus random; dashed: frozen HNSW.}}
\label{fig:q_vs_random}
\end{figure}

\paragraph{Are the learned values needed?}
\label{sec:eval-learning}
The candidate pool already carries structural information without any
learning, so we hold navigation (UCB(64)), binary endpoint reward, candidate pool ($B_{nn}$)
and entry point fixed and change only the score that ranks each node's pool at
rewire time: the learned $\Qhat(u,v)$ versus a uniformly \emph{random} order.
The proposal policy is identical for the two rankers; under GT labels, $B_{nn}$ can contain true neighbors, so this comparison tests ranking within the same informed pool. Labels are either exact (GT) or the serving index's
own top-$k$ (self); the endpoint reward is
$r(u,v,q)=\indic[v\in L(q)]$, where $L(q)$ is the selected label set, while
test recall is always scored against exact ground truth.
Figure~\ref{fig:q_vs_random} reports the mean over $3$ seeds after $1000$ training queries.
The learned ranking improves recall@100 over random in all four settings:
by $0.063$ (GT) and $0.072$ (self) on GIST-1M, and by $0.212$ (GT) and
$0.169$ (self) on T2I-10M. With self labels, learned ranking reaches
$0.877$ on GIST-1M and $0.753$ on T2I-10M, compared with random ranking's
$0.805$ and $0.584$ and frozen HNSW's $0.772$ and $0.605$, respectively.
With GT labels, learned ranking reaches $0.893$ and $0.763$; random
ranking reaches $0.830$ and $0.552$.

\subsection{Combined Axes Testing}
\label{sec:eval-combinatorial}
\begin{figure}[t]\centering
\includegraphics[width=\linewidth]{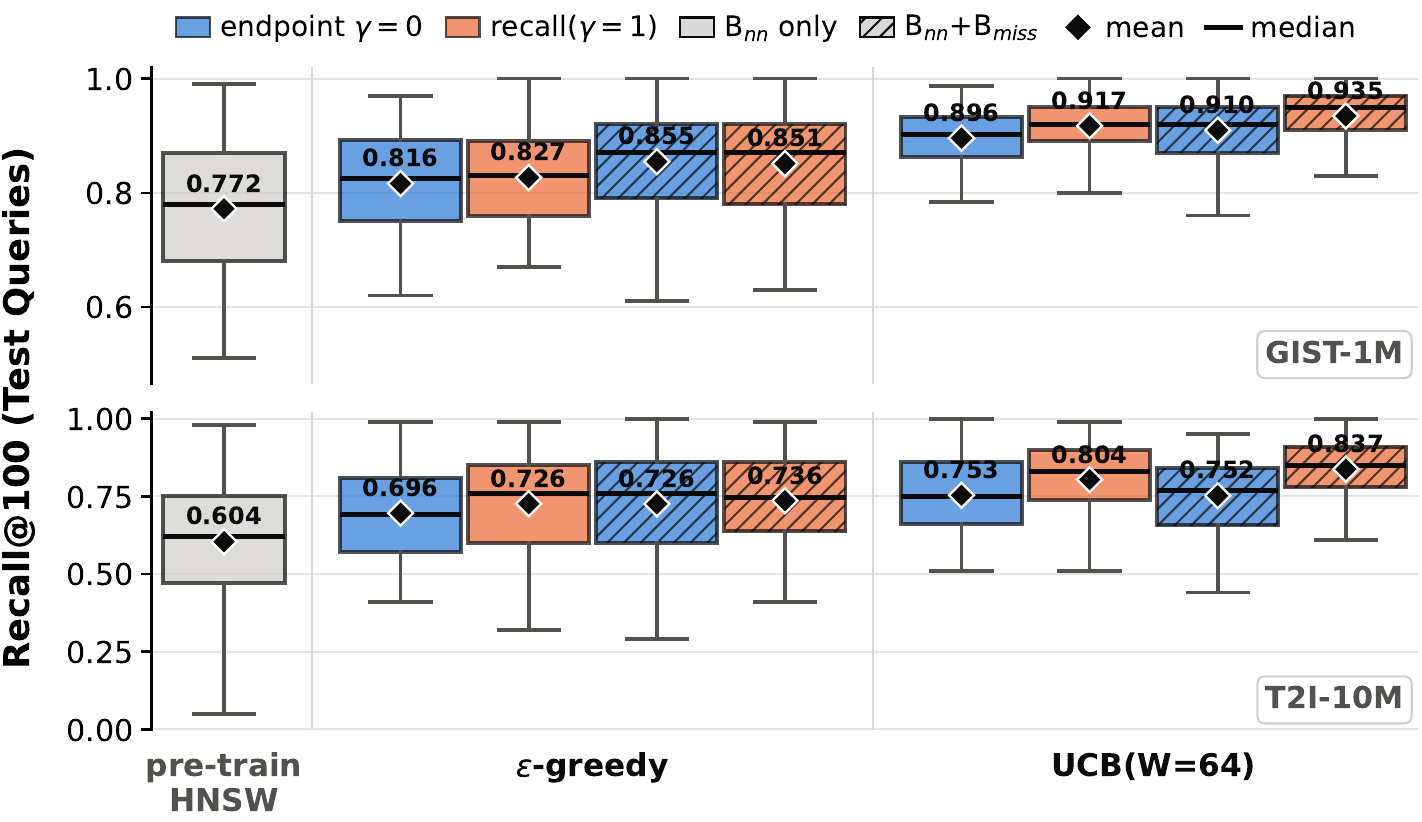}
\caption{Converged recall@100 across navigation, reward and candidate-pool combinations.}
\label{fig:a1_bar}
\end{figure}

We sweep the three design-space axes \emph{jointly} to expose
interactions. The Cartesian product of the reference setting and the best
observed variant from \S\ref{sec:eval-isolated} on each axis
gives $2^3=8$ combined \textcolor{blue}{\textsc{REWEAVE}} variants, reported as eight learned cells in
Figure~\ref{fig:a1_bar}; the boxes show per-query recall distributions at
$k=100$.

On both datasets the dominant factor is navigation: with reward and candidate
pool fixed, switching from $\epsilon$-greedy to UCB(64) lifts mean
held-out recall by $+3$ to $+10$ points across the cells. The other two axes matter conditional on navigation. Under $\epsilon$-greedy,
adding $B_{miss}$ is the larger intervention on GIST-1M, raising mean recall by
$2.5$--$4$ points while the reward choice moves recall by at most a point; on
T2I-10M the two are comparable at $1$--$3$ points each. $B_{miss}$ gain
shrinks under the beam---roughly halving on GIST-1M and surviving only in the
recall-reward cell on T2I-10M---because the wider beam already examines and
rewires toward many of the true neighbors $B_{miss}$ would inject. The reward
axis shows the opposite conditioning: it is amplified under UCB(64), where
recall raises the cells by ${\sim}2.5$ points on GIST-1M and $5$ to
$8.5$ points on T2I-10M. The best cell on both datasets is UCB(64) with recall reward and
$B_{nn}{+}B_{miss}$, which raises mean recall over HNSW by
$16.3$ points on GIST-1M and $23.3$ points on T2I-10M.

\subsection{Locality of Workload}
\label{sec:eval-locality}

\begin{figure}[t]\centering
\includegraphics[width=0.8\linewidth]{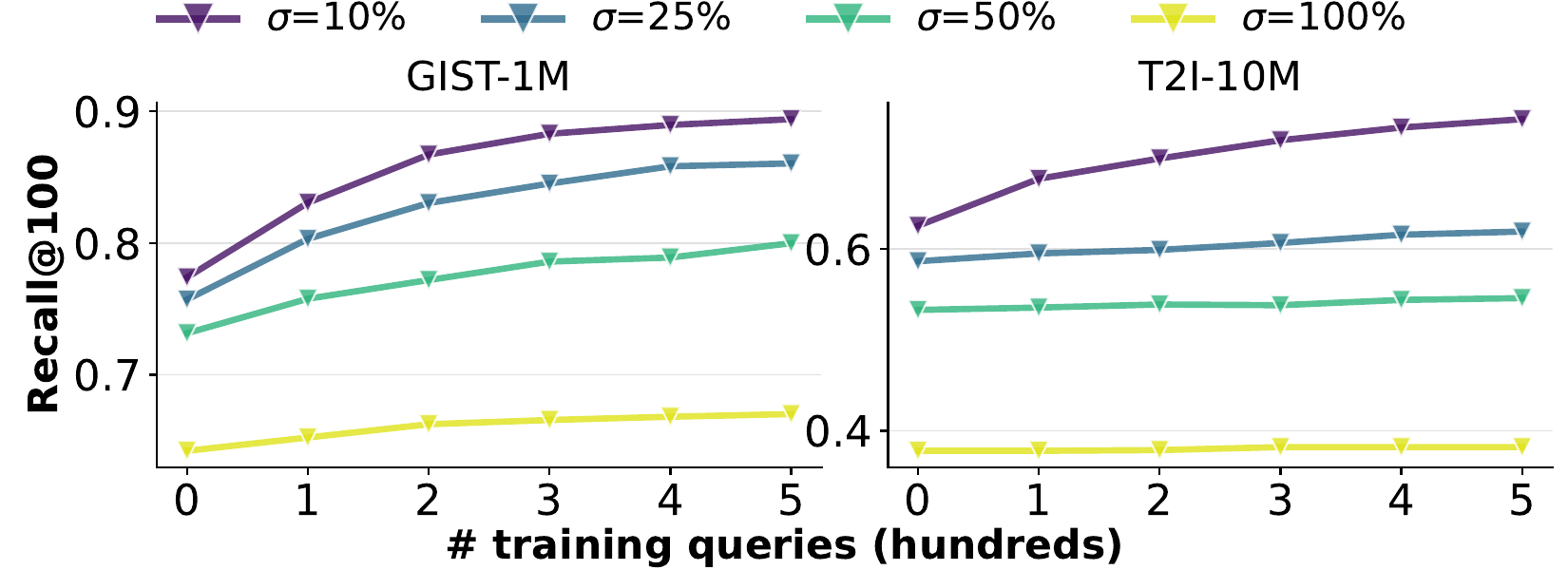}
\caption{Test query recall on different $\sigma$-value tiers.}
\label{fig:locality}
\end{figure}

We vary the Workload~1 noise scale $\sigma\in\{0.10,0.25,0.50,1.00\}$ (in units of $\operatorname{diam}(\mathcal C)$), training the
best variant (bold in Table~\ref{tab:qlearn-design-space}) on $500$ queries per
tier (Figure~\ref{fig:locality}). The lift shrinks monotonically with spread,
from $+0.120$ to $+0.028$ on GIST-1M and to nearly zero on T2I-10M: with less
query overlap, reinforced edges are reused less often. Recall never falls below
the frozen baseline.

\iftechreport
\subsection{Algorithm Convergence}
\label{sec:eval-convergence}

Figure~\ref{fig:f13_diag} tracks value drift, the mean
absolute per-edge change in $\Qhat$ between consecutive rewires, averaged
over all edge and candidate entries scored at both rewires (entries added or
evicted in between are excluded), and edges added per rewire
($B{=}100$) for the best variant. Drift falls several fold after an initial
rise; additions drop from tens of thousands to about a thousand per rewire.
Both curves still decline at the end.

\begin{figure}[t]\centering
\includegraphics[width=0.8\linewidth]{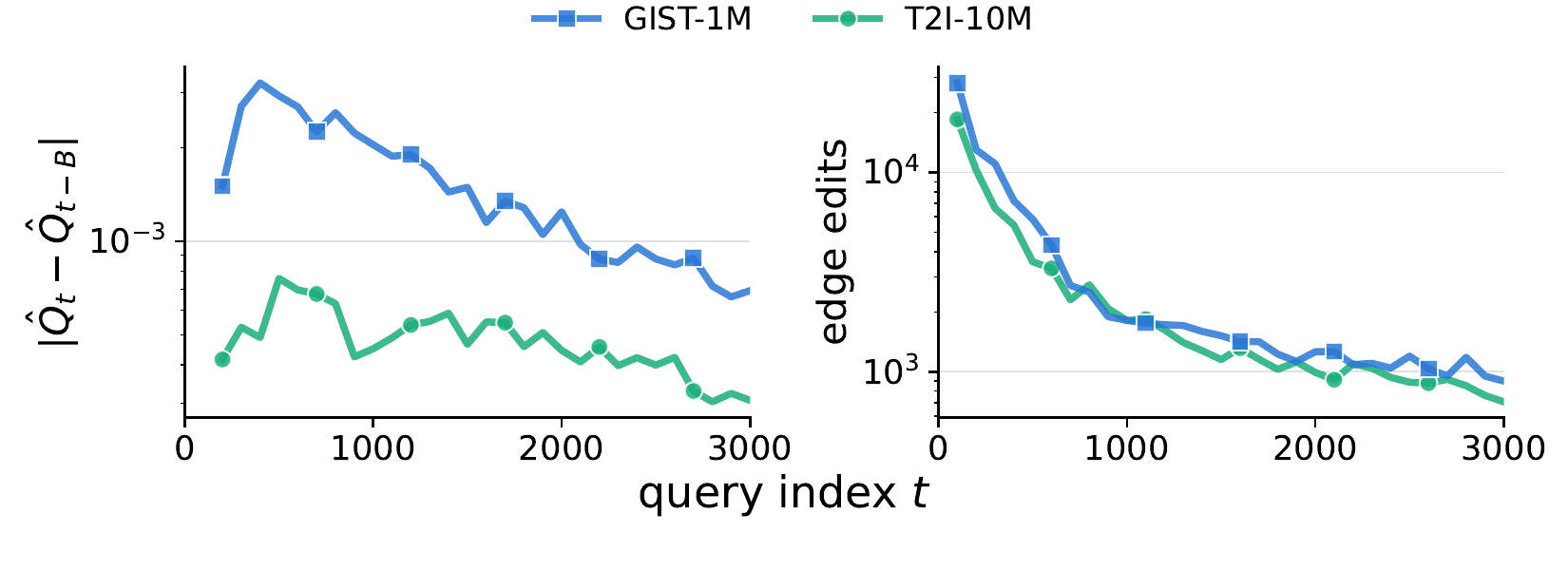}
\caption{Value drift (left) and edge modifications per rewire (right).}
\label{fig:f13_diag}
\end{figure}

\fi

\subsection{Two-rate Feedback Evaluation}
\label{sec:eval-tworate}

\begin{wrapfigure}{R}{0.4\linewidth}\centering
\includegraphics[width=\linewidth]{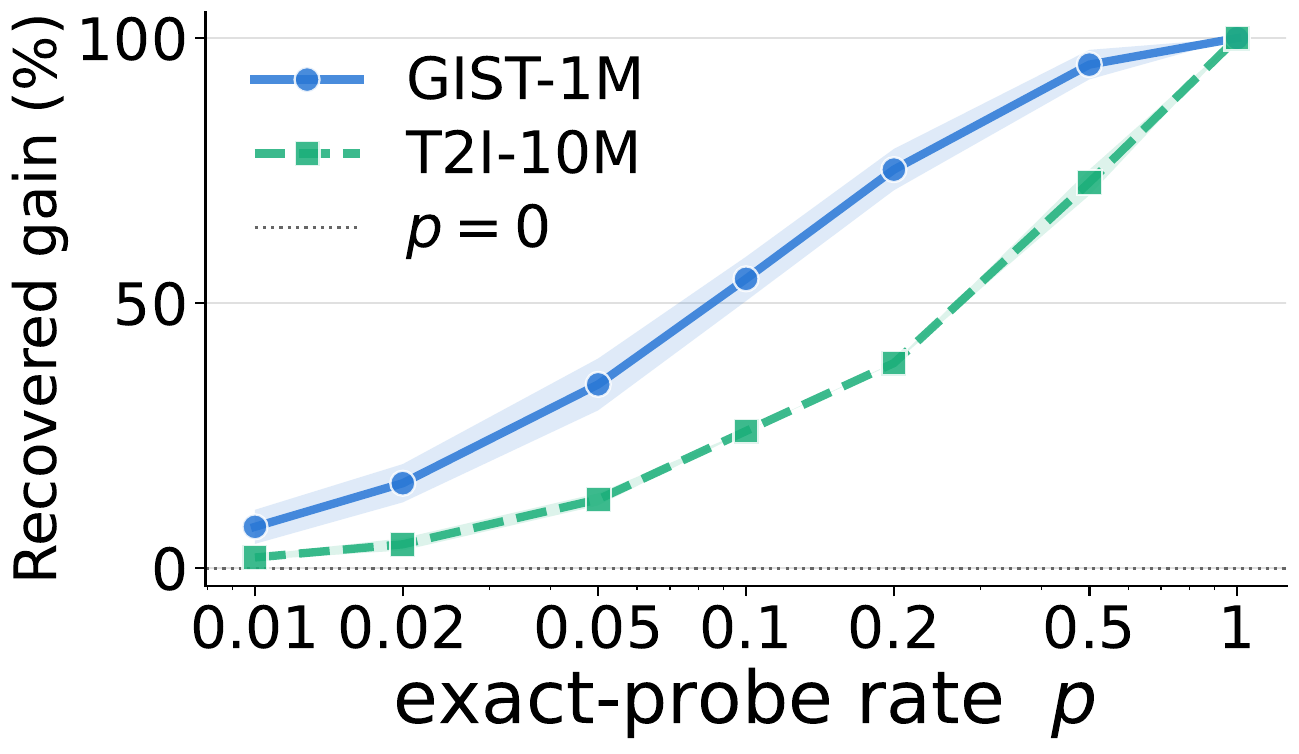}
\caption{Recall gain from exact probes with self-labeled MC($0.8$).}
\label{fig:tworate}
\end{wrapfigure}

We vary the exact-probe rate $p$ while using self-labeled MC($0.8$) on
every query. Exact neighbors populate $B_{miss}$ on an independent
Bernoulli($p$) fraction of queries, isolating the contribution of exact
feedback to candidate proposals.
Let $\mathcal R_s(p)$ denote held-out recall for seed $s$ after $1000$ training queries;
Figure~\ref{fig:tworate} plots
$100\,\frac{\mathcal R_s(p)-\mathcal R_s(0)}{\mathcal R_s(1)-\mathcal R_s(0)}$, the fraction of recall gain
from probing every query ($p=1$) over no exact probes ($p=0$) recovered.
The horizontal axis uses a log scale for $p>0$; the dotted line marks $p=0$.
The denominator is the additional recall obtained by supplying exact
neighbors to $B_{miss}$ on every query, relative to self-feedback alone.
Curves show mean $\pm$ SEM over five GIST-1M seeds and two T2I-10M seeds:
$p{=}0.1$ recovers about $55\%$ of this gain on GIST-1M and $p{=}0.5$
about $73\%$ on T2I-10M; no exact probes gives $0\%$ and probing every
query gives $100\%$ by definition.

\subsection{System Level Comparison}
\label{sec:eval-system}

\textcolor{blue}{We first profile training cost, then compare with
query-aware methods under the adaptation schedules specified below.}

\begin{figure}[t]\centering
\includegraphics[width=0.8\linewidth]{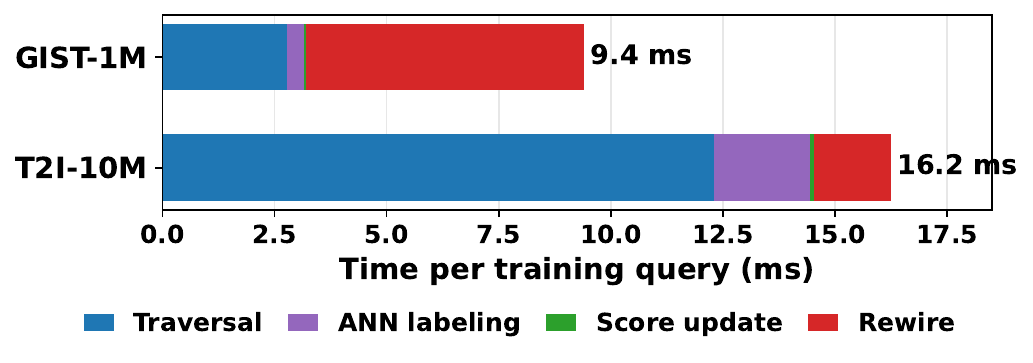}
\caption{\textcolor{blue}{Time per training query of \textsc{REWEAVE} with ANN labels (single thread).}}
\label{fig:train_decomp}
\end{figure}

\paragraph{\textcolor{blue}{Training cost}}
\label{sec:eval-hnsw}

\textcolor{blue}{A single-thread profile}
with ANN labels (\textcolor{blue}{$2$ passes over $1$k training queries};
Figure~\ref{fig:train_decomp}) shows score updates take ${<}1\%$, and time goes to rewiring on GIST-1M ($66\%$,
pruning over $960$-d distances) and traversal on T2I-10M ($76\%$).
Rewiring is per-node, so it parallelizes: on GIST-1M ($20$ rewires) it takes
$19.6$\,s on one thread and $4.5$\,s on $8$, bounded by the serial edge-apply
step ($2.8$\,s); a concurrent search thread loses about half as many queries.

\paragraph{Comparison with query-aware baselines}
\label{sec:eval-roargraph}

\textcolor{blue}{We select as baselines methods that incrementally augment a
graph index to adapt to an out-of-distribution query distribution or to
improve graph search quality.}
NGFix~\cite{ngfix2026} repairs graph defects using query-neighborhood
information and supports online construction. Our implementation applies
its repairs in batches.
Our NAP baseline applies RoarGraph's Neighborhood-Aware
Projection~\cite{chen2024roargraph} as query-neighborhood augmentation of
the fixed HNSW\footnote{RoarGraph implementation: \url{https://github.com/matchyc/RoarGraph}},
linking each past query's nearest neighbors. We evaluate this NAP component,
not the complete RoarGraph index-construction and search pipeline.
We evaluate the full CORAL workflow~\cite{coral}, including coverage-aware
adaptive pruning, fully neighborhood-aware projection, and targeted
connectivity enhancement, updated every $100$ queries.
\textcolor{blue}{Our CleanANN baseline applies CleanANN's guided bridge
building~\cite{cleann2026} to the fixed HNSW after every training query,
linking nodes at selected depths of that query's search tree under
RobustPrune ($\alpha{=}1.2$); we evaluate this component, not CleanANN's
full dynamic-update index.}
All start from the same HNSW ($M_{\rm HNSW}{=}32$; $ef_{\rm construction}{=}200$) and retain the $64$-edge layer-0 cap.
\textcolor{blue}{\textsc{REWEAVE}} uses $W{=}64$, a $64$-entry neighbor
buffer $B_{nn}$ beside the $64$ incumbents ($B_{miss}$ is unused under
self-labels), and
$B{=}100$; training uses $1$k queries.
NGFix, NAP, CORAL and \textcolor{blue}{\textsc{REWEAVE}} use the serving ANN search's returned IDs as
self-labels; CleanANN uses no labels. All five share the same degree cap. \textcolor{blue}{We run this comparison on
GIST-1M, T2I-10M and, to test scale, T2I-100M.} We measure recall@100 against
the exact top-$100$ neighbors and single-thread QPS over
$ef_{\rm search}\in[10,800]$ on a held-out set of $100$ test queries.

\begin{figure}[t]\centering
\includegraphics[width=\linewidth]{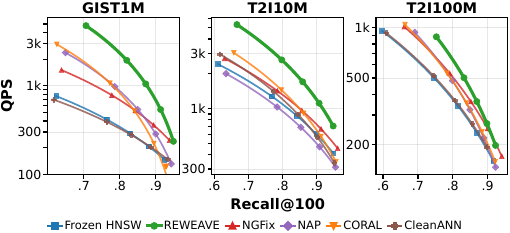}
\caption{\textcolor{blue}{QPS vs.\ recall@100 on $100$ held-out test queries after $1$k self-labeled training queries. NAP: RoarGraph projection; CORAL: full method; CleanANN: guided bridge building.}}
\label{fig:a4_labelfree}
\end{figure}

\textcolor{blue}{\textsc{REWEAVE}} has the highest QPS over nearly the whole recall range
(Figure~\ref{fig:a4_labelfree}): at recall@100 $0.80$, $2.1\times$ NGFix on
T2I-10M and $3.3\times$ on GIST-1M. \textcolor{blue}{On T2I-100M it leads from recall@100
$0.75$ to ${\sim}0.92$, with ${\sim}1.3\times$ NGFix's QPS at $0.80$, and meets
NGFix at the top of its range.} \textcolor{blue}{CleanANN stays close to frozen HNSW: it mainly adapts the
graph to an out-of-distribution query distribution, which need not raise
recall or QPS for a skewed in-distribution workload.} \textcolor{blue}{\textsc{REWEAVE}} also adapts online within the degree
budget without storing queries. In the drift comparison, NGFix, NAP\textcolor{blue}{,}
CORAL \textcolor{blue}{and CleanANN} process new queries in batches during shared adaptation windows
(\S\ref{sec:eval-drift}).

\subsection{ESCI Drift Workload}
\label{sec:eval-drift}

ESCI~\cite{reddy2022shopping} product-search queries are labeled with 14
categories from Amazon's browse-node taxonomy\textcolor{blue}{~\cite{amazoncreatorsapi}}
(keyword and title matching, falling back to the nearest centroid) and embedded
with an ESCI-tuned MiniLM\textcolor{blue}{~\cite{metarankesciminilm} (unit normalized)}.

The stream visits five random categories in order, \textsf{clothing} (A),
\textsf{electronics} (B), \textsf{home\_kitchen} (C), \textsf{books\_media} (D)
and \textsf{baby\_kids} (E), $5$k queries each, then returns to A for $5$k
queries ($30$k total). Within each category cluster, we sample with
replacement from its base-query pool and use the stored embeddings
directly. The returning A segment samples from the same pool as the first A
segment, so \textcolor{blue}{queries} may repeat within and across those segments.
We run three seeds; within a seed every method sees the same stream.

\label{sec:eval-drift-baselines}

\begin{figure}[t]\centering

\includegraphics[width=\linewidth]{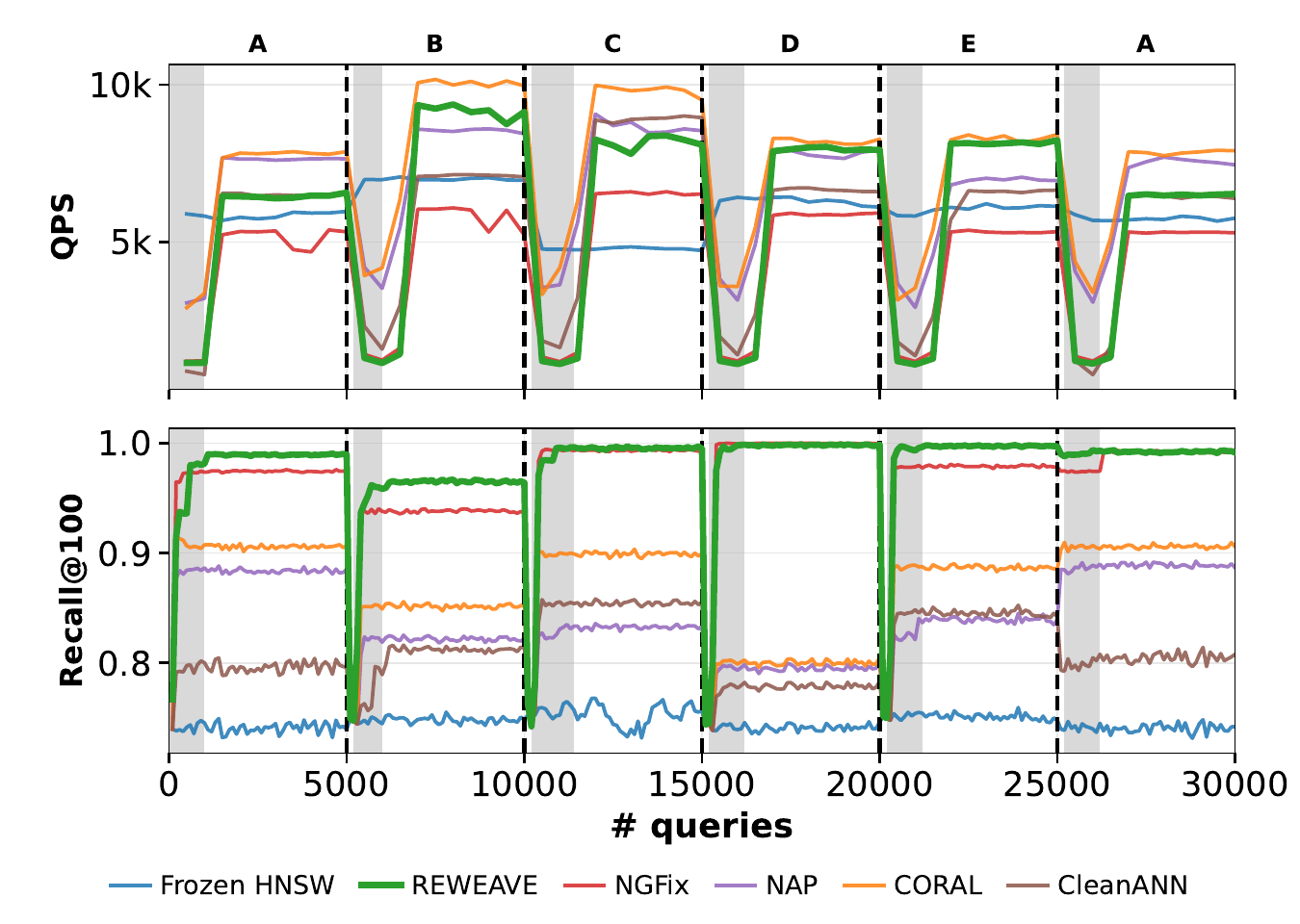}
\caption{\textcolor{blue}{ESCI drift stream, A--E then back to A (seed~1). QPS includes search, adaptation, detector and sync time. Gray: adaptation windows; dashed: category switches.}}
\label{fig:esci_drift_baselines}
\end{figure}

Baselines are frozen HNSW, NGFix, NAP\textcolor{blue}{,} full CORAL \textcolor{blue}{and CleanANN} (\S\ref{sec:eval-roargraph}),
starting from the same HNSW (inner
product, $M_{\rm HNSW}{=}16$, $ef_{\rm construction}{=}200$), searched with
$k{=}ef_{\rm search}{=}100$. \textcolor{blue}{\textsc{REWEAVE}}, NGFix, NAP and CORAL all
use the top-$100$ IDs returned by the serving ANN search as self-labels\textcolor{blue}{, CleanANN uses no labels,}
and \textcolor{blue}{all} share a layer-0 degree cap of $32$. Exact ground truth is used to
evaluate recall.
Let $\operatorname{ANN}_{k_{\rm det}}(q)$ be the detector's $k_{\rm det}{=}100$
returned neighbor IDs. After adaptation, it fixes
$S_{\mathrm{ref}}=\bigcup_{q\in \mathcal W_{\mathrm{ref}}}\operatorname{ANN}_{k_{\rm det}}(q)$
from reference window $\mathcal W_{\mathrm{ref}}$ until the next adaptation ends.
It feeds $z(q)=|\operatorname{ANN}_{k_{\rm det}}(q)\setminus S_{\mathrm{ref}}|/k_{\rm det}$
to ADWIN~\cite{bifet2006adwin}, triggering on increased mean novelty.
All methods share adaptation windows 
ending when the mean absolute edge-value update over a $100$-query window
falls below $0.5$, after which $S_{\mathrm{ref}}$ is refreshed. Baselines
update every $100$ queries in those windows. QPS
includes serving search, self-label processing, adaptation,
detector maintenance and graph synchronization. Baselines replay the shared adaptation windows and share the same detector overhead ($0.047s$ over $30k$ queries) for fair accounting.
Figure~\ref{fig:esci_drift_baselines} shows seed~1 of the three seeds in
Table~\ref{tab:esci_drift_baselines}. QPS uses $500$-query windows;
recall uses $100$-query windows.

\begin{table}[t]\centering\footnotesize
\setlength{\tabcolsep}{3pt}
\begin{tabular}{@{}lrrrrr@{}}
\toprule
Method & Stream R@100 & Return R@100 & Dist./q & Adapt.\,(s) & $\Delta$Mem.\,(MiB) \\
\midrule
Frozen HNSW        & $0.745$ & $0.741$ & $1823$ & -- & -- \\
\textcolor{blue}{\textsc{REWEAVE}} & $\mathbf{0.979}$ & $\mathbf{0.993}$ & $\mathbf{1207}$ & $4.6$ & $+51$ \\
NGFix              & $0.970$ & $0.979$ & $1700$ & $5.3$ & $+10$ \\
NAP                & $0.842$ & $0.887$ & $1452$ & $1.1$ & $+7.2$ \\
CORAL              & $0.875$ & $0.905$ & $1285$ & $1.1$ & $+1.3$ \\
\textcolor{blue}{CleanANN} & \textcolor{blue}{$0.813$} & \textcolor{blue}{$0.802$} & \textcolor{blue}{$1582$} & \textcolor{blue}{$4.2$} & \textcolor{blue}{${\approx}0$} \\
\bottomrule
\end{tabular}
\caption{\textcolor{blue}{ESCI drift stream with a returning segment ($30$k queries, mean of $3$ seeds; std $\le 0.014$). Stream/Return R@100: over all queries / the first $1$k returning A queries. Dist./q: distance computations per query; Adapt.: total adaptation time; $\Delta$Mem.: peak memory minus Frozen HNSW's (${\approx}6.94$\,GiB).}}
\label{tab:esci_drift_baselines}
\end{table}

\textcolor{blue}{\textsc{REWEAVE}} and NGFix lift recall from ${\sim}0.75$ to $0.97$--$1.0$
in each adaptation window (Figure~\ref{fig:esci_drift_baselines},
Table~\ref{tab:esci_drift_baselines}); NAP\textcolor{blue}{,} full CORAL \textcolor{blue}{and CleanANN} achieve stream
recall of $0.842$\textcolor{blue}{,} $0.875$ \textcolor{blue}{and $0.813$}, respectively.
\textcolor{blue}{\textsc{REWEAVE}} needs $34\%$ fewer distance computations than
frozen HNSW and $29\%$ fewer than NGFix. Table~\ref{tab:esci_drift_baselines}
reports total adaptation time: $4.6$\,s for \textcolor{blue}{\textsc{REWEAVE}}
and $5.3$\,s for NGFix.
Peak resident memory is within $51$\,MiB of frozen HNSW
for every method. All methods
keep the degree cap and add at most $0.05\%$ edges.

\paragraph{When to adapt}

Table~\ref{tab:esci_policy} compares four policies: \emph{never},
\emph{always}, the \emph{detector} above, and an \emph{oracle} that starts at
each switch and stops when the recall gain over the last interval falls below
$5\%$ of the gain since the switch. The detector adapts on $19.8\%$ of
queries and comes within $0.008$ stream and $0.003$ return R@100 of always
adapting, at $7.5\times$ its QPS ($1460$ vs.\ $195$). The oracle adapts on
$5.2\%$ of queries and reaches $1.5\times$ the detector's QPS ($2228$),
with stream recall lower by $0.005$ and return recall lower by $0.013$
($1.3$ percentage points). Compared with frozen HNSW, the detector raises
stream R@100 from $0.745$ to $0.979$ and return R@100 from $0.741$ to
$0.993$, while throughput falls from $5830$ to $1460$ QPS, \textcolor{blue}{about}
fourfold.

\begin{table}[t]\centering\footnotesize
\begin{tabular}{@{}lrrrr@{}}
\toprule
Policy & Stream R@100 & Return R@100 & Adapting & QPS \\
\midrule
never    & $0.745$ & $0.741$ & $0\%$     & $5830$ \\
detector & $0.979$ & $0.993$ & $19.8\%$  & $1460$ \\
always   & $0.987$ & $0.996$ & $100\%$   & $195$ \\
oracle   & $0.974$ & $0.980$ & $5.2\%$   & $2228$ \\
\bottomrule
\end{tabular}
\caption{\textcolor{blue}{\textsc{REWEAVE} adaptation policies on the ESCI drift stream; setup and metrics as in Table~\ref{tab:esci_drift_baselines}. Adapting: share of queries adapted on. QPS includes search, adaptation (with label search) and sync time.}}
\label{tab:esci_policy}
\end{table}

\section{Related Work}
\label{sec:related}

HNSW~\cite{malkov2020hnsw}, NSG~\cite{fu2019nsg}, and disk/memory hybrids such as DiskANN~\cite{subramanya2019diskann}, SPANN~\cite{chen2021spann} and HM-ANN~\cite{ren2020hmann} are the dominant vector search
indices; surveys confirm their lead on the recall--latency
frontier~\cite{wang2021survey, ma2026comprehensivesurveyvectordatabase}. All build the graph from the \emph{data} alone
and do not revisit it for the query workload. Learned
indices~\cite{kraska2018learnedindex,ding2020alex,ferragina2020pgm} apply
learning to other access patterns.

Workload-adaptive databases have transformed modern data management~\cite{chaudhuri2007selftuning,idreos2011merging,petraki2015holistic}. These include cost-driven index selection and "what-if" optimization utilities~\cite{chaudhuri1997efficient,chaudhuri1998autoadmin,wang2026evaluating}.

Methods that tune ANN search behavior~\cite{ li2020adaptive_termination, oguri2024entrypoint} modify traversal path rather than the topology. Several query-aware approaches build graphs from sampled queries before deployment~\cite{jaiswal2022ooddiskann, huang2015qalsh, chen2024roargraph, ruan2025gate};
query-hardness measures~\cite{wang2024steinerhardness} guide such construction.
\textcolor{blue}{OOD-DiskANN~\cite{jaiswal2022ooddiskann} and CDMG~\cite{yue2026cdmg} adapt an
entire index to an out-of-distribution query workload, but build it as a
whole rather than updating the graph incrementally as queries arrive.}

NGFix~\cite{ngfix2026} guides graph repairs around difficult
queries and supports online graph construction.
CORAL~\cite{coral} incrementally constructs cross-modal graphs \textcolor{blue}{via}
coverage-aware adaptive pruning, Fully Neighborhood-Aware Projection
(FNAP), and targeted connectivity enhancement. \textcolor{blue}{\textsc{REWEAVE}} learns
edge values from exact/self feedback to select candidates within a fixed
degree budget.
Streaming indices~\cite{singh2021freshdiskann,xu2025inplace, yu2025topologyaware, liu2025wolverine} maintain the graph
under \emph{data} insertion and deletion, an orthogonal form of dynamism.
Filtered ANN~\cite{gollapudi2023filtereddiskann,zuo2023arkgraph,peng2025dynamic} adapts to predicate
constraints\textcolor{blue}{, not} query workloads.

Reinforcement learning has been applied at the retrieval application layer
(query reformulation~\cite{nogueira2017queryreformulation}); routing on a
proximity graph has been learned for
recommendation~\cite{feng2022reinforcementrouting} and made
probabilistic~\cite{lu2024probabilisticrouting}, but these choose among
\emph{existing} edges rather than re-allocating the degree budget, and offer no
regret guarantee. Meanwhile, we utilize stochastic approximation~\cite{robbins1951stochastic,
kushner2003stochastic}, finite-armed bandits and the UCB1
analysis~\cite{lai1985allocation,auer2002ucb1,lattimore2020bandit}, sleeping
bandits, where only a changing subset of choices is available at each
decision~\cite{kleinberg2010sleeping}, and
switching/non-stationary bandits~\cite{garivier2011swucb,besbes2014nonstationary,
auer2019adswitch}.

\section{Conclusion}
\label{sec:conclusion}

Under retained evidence, conditional-mean control, and candidate coverage, the learned graph stabilizes at a reward-specific value-greedy graph. Our regret bounds distinguish routing error from graph mismatch. Mixed-feedback estimation further gives sufficient bias and margin conditions for agreement with the exact-reference graph.
Future work will focus on extending this framework for disk resident indexes which raise important challenges for IO optimizations.

\bibliographystyle{ACM-Reference-Format}
\bibliography{qlearn_draft}

\iftechreport

\appendix

\numberwithin{theorem}{section}
\numberwithin{lemma}{section}
\numberwithin{proposition}{section}
\numberwithin{corollary}{section}
\numberwithin{remark}{section}

\section{Observation Model, Estimation, and Stabilization}
\label{app:proofs8}\label{app:prelim}
We use the definitions and assumptions of \S\ref{sec:analysis}. In
particular, $K$ counts fixed node--candidate pairs over all possible decision
nodes, not the size of one instantaneous pool. In particular,
$K\le |V_\circ|C$ holds only when every entire universe has size at most
$C$, rather than merely each current pool. All probabilities include
query and navigation randomness. The comparison graph and policy are
$(G^\dagger,\pi^\dagger)$ for the selected credit rule. Rewards used for
estimation are denoted $Y_t(e)$, the Step 2 observations specified in
\S\ref{sec:estimator}; collected endpoint reward is $R_t$.
The proofs below do not equate either quantity with returned-set recall.

\subsection{A common concentration argument}
\label{app:concentration}
Fix an edge $e$ and suppress it in this subsection. Let $I_t$ indicate its
evaluation on query $t$, and let
$\mu_t=\mathbb E[Y_t\mid\mathcal F_{t-1},I_t=1]$ whenever that event has
positive probability. Give $\mu_t$ any value in $[0,1]$ otherwise. This
conditional expectation is a function of the preceding history; it
integrates over the current query and the selection event. The indicator
$I_t$ need not be independent of the reward. Since the reward lies in
$[0,1]$, conditional Hoeffding's lemma, first given $I_t$ and then given
$\mathcal F_{t-1}$, gives for every real $z$
\begin{equation}
 \mathbb E\!\left[
 e^{z I_t(Y_t-\mu_t)-z^2I_t/8}\mid\mathcal F_{t-1}
 \right]\le1.
 \label{eq:conditional-mgf}
\end{equation}
On $I_t=0$ the integrand is $1$; on $I_t=1$ its conditional expectation is
at most $1$. Consequently
\[
 M_t(z)=\exp\!\left(z\sum_{s\le t}I_s(Y_s-\mu_s)-z^2N_t/8\right)
\]
is a nonnegative supermartingale starting at $1$.

At the $m$th evaluation, the usual exponential bound (or optional stopping
of $M_t$) yields
\begin{equation}
 \Pr\!\left(\left|\frac1m\sum_{j=1}^m(Y^{(j)}-\mu^{(j)})\right|\ge x,
             \ m\text{ evaluations occur}\right)
 \le2e^{-2mx^2}.
 \label{eq:sample-tail}
\end{equation}
The qualification avoids conditioning the probability on a random sampling
event. When evaluations are infinite almost surely it can be omitted.
We will also use the following bound for random sample counts:
\begin{equation}
 \Pr\!\left(\exists t:N_t\ge m,
       |\widehat Q_t-\theta_r|>\kappa+x\right)\le2e^{-2mx^2}.
 \label{eq:anytime-slope}
\end{equation}
Indeed, Assumption~\ref{ass:means} bounds the average conditional mean
within $\kappa$ of $\theta_r$. On a positive deviation in
\eqref{eq:anytime-slope}, use $z=4x$: then $M_t(z)\ge e^{2N_tx^2}\ge
e^{2mx^2}$. The maximal inequality for a nonnegative supermartingale bounds
this event by $e^{-2mx^2}$. Apply the same argument with $z=-4x$ and add.
This proof permits adaptive graph changes, stopping, and observation times;
it does not substitute a random count into a fixed-count inequality.
Under Assumption~\ref{ass:means} and the set margin of
Assumption~\ref{ass:margins}, with retained running means, fix a failure
probability $\delta\in(0,1)$ and put
\begin{equation}
 m_{\rm set}=\left\lceil\frac{8}{g_{\rm set}^2}
                   \log\frac{2K}{\delta}\right\rceil.
 \label{eq:sample-budget}
\end{equation}
Taking $x=g_{\rm set}/4$ and $m=m_{\rm set}$ in
\eqref{eq:anytime-slope}, then unioning over $K$ pairs, gives the following
simultaneous guarantee: with probability at least $1-\delta$, every pair's
score is within $\kappa+g_{\rm set}/4$ of its reference whenever its count
is at least $m_{\rm set}$. On this event, a batch with that many observations
of every compared pair selects $G^\dagger$ whenever all target edges are
available. The event is unconditional; it does not condition on the
counts reaching the threshold. This is an observation requirement per
pair; the queries needed to obtain those observations depend on coverage.
Halving the effective margin multiplies the sufficient evidence by
roughly four. The unknown reference margins and conditional-mean
tolerance prevent treating this bound as an empirical stopping certificate.

For completeness, if
$\Pr(I_t=1\mid\mathcal F_{t-1})\ge p_t$ for deterministic $p_t\in[0,1]$,
then, with $A_t=\sum_{s\le t}p_s$,
\begin{equation}
 \Pr(N_t<A_t/2)\le e^{-A_t/8}.
 \label{eq:count-tail}
\end{equation}
To see this, for $z>0$ condition successively on the history to obtain
$\mathbb E e^{-zN_t}\le\exp(-(1-e^{-z})A_t)$.
Markov's inequality with $z=\log2$ proves a stronger exponent than
$-A_t/8$. No independence between successive evaluation events is needed.

\subsection{Evaluation coverage}
\label{app:lem1}
\begin{lemma}[Evaluation coverage]
\label{lem:infinite}
An edge is evaluated infinitely often almost surely on the event
$\sum_t\Pr(I_t(e)=1\mid\mathcal F_{t-1})=\infty$.
For single-path $\epsilon$-greedy, the evaluation probability is at least
$\epsilon_t\beta_t(e)/M$, where $\beta_t(e)$ is the conditional probability
that the current out-edge leads from a visited source node to a destination
node closer to the query, before the episode cap is reached.
\end{lemma}
\begin{proof}
The conditional Borel--Cantelli lemma gives
\[
 \begin{aligned}
 &\{I_t(e)=1\text{ infinitely often}\}\\
 &\quad=\left\{\sum_t\Pr(I_t(e)=1\mid\mathcal F_{t-1})=\infty\right\}
 \quad\text{a.s.}
 \end{aligned}
\]
For an improving current out-edge at a visited source node before the cap, the
single-path exploration coin is independent of the preceding route.
A uniform permutation puts that edge first among the improving edges with
probability at least $1/M$. Integrating over the event defining
$\beta_t(e)$ gives the stated lower bound. It only applies while the edge
is a current out-edge. Neither absolute continuity of the workload nor
positive visitation at each finite time supplies a uniform lower bound on
$\beta_t(e)$.

The distinction matters even on a fixed graph. Let the entry be $10$,
with chain $10\to9\to8\to7$ and an additional exit to $0$ at each chain
node, and draw $q$ uniformly near $0$, with $k=1$. Endpoint exploitation
always prefers the exit, including initially by distance tie-breaking.
Reaching $7$ requires three exploratory choices and has probability
$(\epsilon_t/2)^3=\Theta(t^{-3/2})$. Its sum is finite, so $7$ is visited
only finitely often almost surely. This is why coverage is an explicit
condition, rather than a consequence of a finite graph or a decaying
exploration schedule.
\end{proof}

\subsection{Value estimation}
\label{app:thm1}
\begin{theorem}[Value estimation]
\label{thm:convergence}
Under Assumption~\ref{ass:means}, an infinitely evaluated edge whose
running-mean evidence is retained satisfies
$\limsup_t|\widehat Q_t(e)-\theta_r(e)|\le\kappa$ almost surely.
For fixed $m\ge1$ and $\delta\in(0,1)$, the probability that its $m$th
observation occurs and the resulting error exceeds
$\kappa+\sqrt{\log(2/\delta)/(2m)}$ is at most $\delta$.
The statements apply to endpoint and recall credit with their respective
conditional means; exact convergence is guaranteed when $\kappa=0$.
\end{theorem}
\begin{proof}
The update in \eqref{eq:rm} uses step size $1/m$ on the $m$th recorded
observation. Induction gives $\widehat Q=m^{-1}\sum_{j\le m}Y^{(j)}$;
the first update removes the unobserved prior. Taking
$x=\sqrt{\log(2/\delta)/(2m)}$ in \eqref{eq:sample-tail} proves the
finite-sample assertion. For each rational $x>0$, the tail is summable in
$m$, so Borel--Cantelli gives
$m^{-1}\sum_{j\le m}(Y^{(j)}-\mu^{(j)})\to0$ almost surely.
The average of the conditional means stays within $\kappa$ of $\theta_r$.
On an infinitely evaluated edge, $N_t\to\infty$, proving the limsup bound.
No assertion of independence of the recorded rewards is needed, even when
$\kappa=0$. The argument applies equally to bounded endpoint and recall
credit when their own conditional means satisfy the assumption.
\end{proof}

\subsection{Graph stabilization}
\label{app:prop1}
\begin{proposition}[Graph stabilization]
\label{prop:stable}\label{prop:stabilization}
Under Assumption~\ref{ass:means}, the set margin of
Assumption~\ref{ass:margins}, and Assumption~\ref{ass:coverage}, batched
top-$M$ rewiring eventually selects $G^\dagger$ permanently, almost surely.
If all target edges are available at every batch and every compared edge
has conditional evaluation probability at least $\nu>0$ on every query,
then the expected total number of queries on a different graph is
$O(B+K/(\nu g_{\rm set}^2))$.
\end{proposition}
\begin{proof}
There are $K<\infty$ compared pairs. Under
Assumption~\ref{ass:coverage}, Theorem~\ref{thm:convergence} implies that,
almost surely, after a finite time all their errors are smaller than
$\kappa+g_{\rm set}/4$. If $a$ belongs to the reference top-$M$ set at a
node and $b$ does not, then
\[
 \widehat Q(a)-\widehat Q(b)
 >\Delta_{\rm set}-2\kappa-g_{\rm set}/2=g_{\rm set}/2>0.
\]
At the next sufficiently late batch all target edges are available, so the
selection is exactly the reference set. Incumbents remain in the pool,
and subsequent estimates preserve this ordering. If the universe has at
most $M$ entries, availability alone ensures all of them are retained.
This proves stabilization without a graph-dependent best-response map.

For the quantitative assertion, assume every target is available at every
batch and every compared pair has evaluation probability at least $\nu$.
At query $t$, \eqref{eq:count-tail} gives
$\Pr(N_t(e)<\nu t/2)\le e^{-\nu t/8}$. Combining this with
\eqref{eq:anytime-slope}, with $x=g_{\rm set}/4$, gives
\[
 \Pr\bigl(|\widehat Q_t(e)-\theta_r(e)|>
               \kappa+g_{\rm set}/4\bigr)
 \le e^{-\nu t/8}+2e^{-\nu t g_{\rm set}^2/16}.
\]
A union bound over $K$ edges bounds an incorrect rewire at $t=jB$.
Charge the first $B$ queries directly, and at most $B$ subsequent queries
to each incorrect batch. Since $\sum_{j\ge1}e^{-ajB}\le1/(aB)$,
\begin{equation}
 D_\infty\le B+\frac{8K}{\nu}+\frac{32K}{\nu g_{\rm set}^2}
 =O\!\left(B+\frac{K}{\nu g_{\rm set}^2}\right).
 \label{eq:structural-cost}
\end{equation}
The formula uses $g_{\rm set}\le1$, as may be assumed for rewards in
$[0,1]$. Universes without a selection boundary need no estimation test.
This bounds the expected number of incorrect-graph queries directly; it
does not infer an expected stabilization time from almost-sure convergence.
The hypotheses can be supplied by universe-scoring with uniform visitation
to source nodes or by additional probes, whose observation and computational
costs are part of the feedback regime. Traversal-only sampling of a
permanently fixed graph cannot satisfy infinite coverage of excluded edges.
\end{proof}

\begin{remark}[Reverse proposals]
\label{rem:reverse-proposals}
Proposition~\ref{prop:stable} also applies when selecting $(u,v)$ proposes
$(v,u)$, provided the fixed finite universes include every possible directed
proposal and the recipient uses its own score, pool cap, and top-$M$
selection. The proof depends on accurate ordering and target availability,
so this proposal source leaves the argument unchanged under the same
mean, margin, retained-evidence, and coverage conditions. Reverse proposals
do not themselves supply coverage or a faster rate. Each configuration
uses its own references and comparison graph if these change with the
proposal process. Forcing reciprocal edges after selection can displace
reference edges and is outside the guarantee.
\end{remark}

\section{Stationary Routing Regret}
\label{app:regret}
All stationary routing bounds assume Assumption~\ref{ass:means} and the
navigation margin of Assumption~\ref{ass:margins}, with complete
running-mean evidence retained. The set margin is used when applying
\eqref{eq:structural-cost}. We retain the graph-mismatch term throughout.

\subsection{Coupling and the regret decomposition}
\label{app:coupling}
Run the comparator on the learner's current query, with the same entry
rule and episode cap, and write $R_t^\dagger=\rho(q_t;G^\dagger,\pi^\dagger)$.
Its fixed graph and policy give $\mathbb E[R_t^\dagger]=R^\dagger$
regardless of the learner's history. On $G_t=G^\dagger$, the accepted paths agree until a first different
choice. If no difference occurs, collected endpoint rewards are identical.
Otherwise their difference in absolute value is at most $H$.
Thus
\[
 \left|\mathbb E\sum_{t\le T}(R_t^\dagger-R_t)\right|
 \le\sum_{t\le T}\mathbb E|R_t^\dagger-R_t|
 \le HD_T+H\mathbb E[J_T],
\]
which proves \eqref{eq:regdecomp}. A first difference on the correct graph
is a choice of a strictly lower-reference improving edge, by the navigation
margin. Charge only this first difference in each query. This avoids
assuming a local immediate-reward gap also bounds downstream loss. It also
counts every node where either policy can choose an outgoing edge, not just an intersection of visitation
supports. Equality of graphs here and in $D_T$ concerns their out-lists on
$V_\circ$; rows that can never be used by either policy are irrelevant.

\subsection{Full-information routing}
\label{app:thm2}
\begin{theorem}[Full information]
\label{thm:fullinfo}\label{thm:full_info}
With endpoint universe-scoring at every visited source node and greedy navigation,
$|\mathrm{Reg}(T)|\le HD_T+O(K/g_{\rm nav}^2)$.
Under the quantitative coverage condition of Proposition~\ref{prop:stable},
this is bounded independently of $T$ by
$O(B+K/(\nu g_{\rm set}^2)+K/g_{\rm nav}^2)$.
\end{theorem}
\begin{proof}
Fix a node $u$. Under universe-scoring, each $e=(u,v)$ in its universe has
exactly $m$ observations after $m$ visits to $u$, irrespective of whether
the edge was installed. If all its scores are within
$\kappa+g_{\rm nav}/4$ of their references, greedy ranking chooses the
reference-best improving edge whenever the graph is correct. The initial visit is charged directly. For $m\ge1$, before the
$(m+1)$st visit, the probability of a score error large enough to permit
a different choice is at most
$2|\mathcal U(u)|e^{-m g_{\rm nav}^2/8}$ by
\eqref{eq:sample-tail}. The bound is unconditional and includes the event
that this visit occurs. Each $m$ is charged at most once at this node.
Summing over visits and nodes gives
\[
 \mathbb E[J_T]\le
 \sum_u\frac{2|\mathcal U(u)|}{1-e^{-g_{\rm nav}^2/8}}
 \le2K(1+8/g_{\rm nav}^2).
\]
Use \eqref{eq:regdecomp} for the first assertion and
\eqref{eq:structural-cost} for the second. Scoring only current out-edges
does not give equal sample counts for all compared candidates and is not
the feedback regime of this theorem.
\end{proof}

\subsection{\texorpdfstring{$\epsilon$}{epsilon}-greedy routing}
\label{app:thm3}
\begin{theorem}[$\epsilon$-greedy]
\label{thm:sqrt}\label{thm:eps_greedy}
Let $\epsilon_t=\min\{1,c_\epsilon/\sqrt t\}$. Suppose for constants
$a,A,d>0$ and every reference out-edge $e$,
$\Pr(N_{t-1}(e)<a\sqrt t)\le A e^{-d\sqrt t}$ for all sufficiently large $t$.
Then
$|\mathrm{Reg}(T)|\le HD_T+2c_\epsilon H^2\sqrt T+O(1)$,
where the constant depends on $K,a,A,d,g_{\rm nav}$ and the initial prefix.
On a fixed benchmark graph, a uniform positive lower bound on $\beta_t(e)$
in Lemma~\ref{lem:infinite} suffices for this sample-count condition.
\end{theorem}
\begin{proof}
Each query has at most $H$ navigation choices. The probability that any
exploration coin fires is at most $H\epsilon_t$; charging $H$ for the
resulting trajectory difference yields
\[
 H^2\sum_{t\le T}\epsilon_t
 \le2c_\epsilon H^2\sqrt T.
\]
On the correct graph with no exploration, a different choice requires
an inaccurate score of a reference out-edge. For each such edge, the
sample-count hypothesis and \eqref{eq:anytime-slope} bound its probability
by
\[
 A e^{-d\sqrt t}+2e^{-a g_{\rm nav}^2\sqrt t/8},
\]
apart from the stated finite initial prefix. Both series are summable:
for $c>0$, $\sum_{t\ge1}e^{-c\sqrt t}\le1+2/c^2$.
Union over at most $K$ reference out-edges and charge $H$ per different
query. This gives the finite identification term in the theorem. In
particular, the proof does not bound exploitation mistakes by samples of
the suboptimal edge alone; both compared scores must be accurate.

On a fixed $G^\dagger$, if $\beta_t(e)\ge\beta>0$ for each reference edge,
Lemma~\ref{lem:infinite} implies a deterministic cumulative lower bound
$\sum_{s<t}\Pr(I_s(e)=1\mid\mathcal F_{s-1})\ge c_0\sqrt t$ for all
sufficiently large $t$, where $c_0>0$ depends on $\beta,c_\epsilon,M$.
Applying \eqref{eq:count-tail} with these deterministic per-query lower
bounds verifies the sample-count hypothesis. Uniform eligibility is an
additional geometric/visitation condition, not implied by the workload
having a density.
\end{proof}

\subsection{Proof of Theorem~\ref{thm:ucb}}
\label{app:thm4}
For $t\ge2$, consider the event that every observed edge estimate satisfies
\begin{equation}
 |\widehat Q_{t-1}(e)-\theta_r(e)|
 \le\kappa+\sqrt{2\log t/N_{t-1}(e)}.
 \label{eq:ucb-good}
\end{equation}
For fixed observation count $m\le t-1$, \eqref{eq:sample-tail} bounds its
failure for one edge by $2t^{-4}$. Union over counts and edges gives
failure probability at most $2Kt^{-3}$. Hence the expected number of
queries on which \eqref{eq:ucb-good} fails is $O(K)$.

On a correct-graph query satisfying \eqref{eq:ucb-good}, let $v$ be the
first wrongly selected destination node and $w$ the reference-best improving neighbor of
the common source node. If $w$ has positive count, its index is at least
$\theta_r(u,w)-\kappa$; at zero count its index is infinite. If
\[
 N_{t-1}(u,v)\ge 32\log(T\vee2)/g_{\rm nav}^2,
\]
then its index is at most
$\theta_r(u,v)+\kappa+g_{\rm nav}/2$, strictly smaller than the lower
bound for $w$. Therefore every charged wrong choice on a good query has
count below this threshold. Each such choice increments that edge's count,
so at most the threshold rounded up can be charged to a given edge over
the entire horizon. Counts acquired on other queries only help. Summing
over $K$ edges and adding the $O(K)$ bad-confidence queries proves
\[
 \mathbb E[J_T]=O(K\log(T\vee2)/g_{\rm nav}^2).
\]
Apply \eqref{eq:regdecomp}. The comparison is with the best
\emph{currently improving} edge on the reference graph. A gap from the
maximum over the entire pool would not justify the index comparison.
The argument is an explicit adaptation of the UCB counting method
\cite{auer2002ucb1,kleinberg2010sleeping}; it does not assume the rewards
or their availability are independent across queries.

\subsection{Average collected reward}
\label{app:cor1}
\begin{corollary}[Average collected reward]
\label{cor:average_reward}
If $D_T=o(T)$ and the relevant routing theorem applies, then
$T^{-1}\sum_{t\le T}\mathbb E[R_t]\to R^\dagger$.
Its absolute error is bounded by the corresponding theorem divided by $T$.
For $D_T=O(1)$ the rates are $O(1/T)$, $O(T^{-1/2})$, and $O(\log T/T)$.
\end{corollary}
\begin{proof}
By definition,
\[
 \left|T^{-1}\sum_{t\le T}\mathbb E[R_t]-R^\dagger\right|
 =|\mathrm{Reg}(T)|/T.
\]
The absolute bounds above and $D_T=o(T)$ imply convergence; inserting
$D_T=O(1)$ gives the three displayed rates. No sign assumption about regret
is needed. Without a rate for $D_T$, an almost-sure eventual graph match
implies $D_T/T\to0$ by bounded convergence and Ces\`aro averaging, but
does not supply the faster rates.
\end{proof}

\section{Discounted Estimation and Dynamic Regret}
\label{app:thm5}
Let $\Upsilon_T$ count distribution changes through query $T$, and $i(t)$
denote query $t$'s phase. Phase $i$ has references $\theta_{r,i}$,
comparator $(G_i^\dagger,\pi_i^\dagger)$, and reward $R_i^\dagger$. Define
$\mathrm{Reg}_d(T)=\sum_{t=1}^T(R_{i(t)}^\dagger-\mathbb E[R_t])$ and
$D_T^d=\sum_{t=1}^T\Pr(G_t\ne G_{i(t)}^\dagger)$.
Discounted statistics are retained across rewires and evictions.

For endpoint or recall rewards in $[0,1]$, use the discounted estimates
and observation counts of \S\ref{sec:estimator} to define the index
\begin{equation}
 U_t^\lambda(e)=\widehat Q_{t-1}^\lambda(e)
 +\sqrt{8\ell_T/N_{t-1}^\lambda(e)},\qquad
 \ell_T=\log(2KT^3),
 \label{eq:discounted-ucb}
\end{equation}
with index $+\infty$ at zero count. Here $T\ge2$ is the planned query
horizon. The theorem below specifies the range of the forgetting factor
used by the analysis.

\begin{theorem}[Discounted UCB]
\label{thm:drift}
Suppose Assumption~\ref{ass:means} holds within each phase, with common
$\kappa$ and finite universes, and the navigation margin of
Assumption~\ref{ass:margins} gives a uniform $g_{\rm nav}>0$. For $T\ge2$, use
$\lambda=1-\eta$, $1/T\le\eta\le1/2$, and the discounted-UCB index
\eqref{eq:discounted-ucb}. Then
\[
 |\mathrm{Reg}_d(T)|\le HD_T^d+
 O\!\left(\frac{K(T\eta+1)\ell_T}{g_{\rm nav}^2}
       +\frac{\Upsilon_T\log(8/(g_{\rm nav}\eta))}{\eta}\right).
\]
With a supplied change budget, choosing $\eta=\min\{1/2,\sqrt{(\Upsilon_T+1)/T}\}$ gives a
$\widetilde O(\sqrt{(\Upsilon_T+1)T})$ routing term for fixed $K,g_{\rm nav}$.
Here $\widetilde O$ suppresses logarithmic factors.
No minimum phase length is required. If there are no changes, ordinary UCB
retains the stronger stationary logarithmic bound.
\end{theorem}

We prove Theorem~\ref{thm:drift} using the zero initialization of
\S\ref{sec:estimator}. Write
$\eta=1-\lambda$, $g=g_{\rm nav}$, and $\ell=\ell_T$.
Each edge records at most one observation per query, so
$N_{t-1}^\lambda(e)\le1/\eta\le T$.

\subsection{Weighted concentration}
\label{app:weighted}
Fix a deterministic query index $t$ and an edge $e$. For $s<t$ put
$w_s=\lambda^{t-1-s}$ and
\[
 \begin{gathered}
 Z=\sum_{s<t}w_sI_s(e)(Y_s(e)-\mu_s(e)),\\
 N=\sum_{s<t}w_sI_s(e),\qquad V=\sum_{s<t}w_s^2I_s(e)\le N.
 \end{gathered}
\]
Applying \eqref{eq:conditional-mgf} with deterministic multiplier $w_s$
at query $s$ yields $\mathbb E\exp(zZ-z^2V/8)\le1$.
Peel $1\le N\le T$ into bins $v\le N<2v$, $v=1,2,4,\ldots$.
On a bin, the event $Z\ge\sqrt{8\ell N}$ implies
$Z\ge x=\sqrt{8\ell v}$ and $V<2v$. Markov's inequality with
$z=2x/v$ bounds its probability by
$\exp(-zx+z^2(2v)/8)=e^{-8\ell}$.
The same holds for $-Z$. Union over bins, $K$ pairs, and at most $T$
query indices shows that, except on an event of probability at most
$T^{-2}$,
\begin{equation}
 \left|\widehat Q_{t-1}^\lambda(e)
       -\frac{\sum_{s<t}w_sI_s(e)\mu_s(e)}{N_{t-1}^\lambda(e)}\right|
 \le\sqrt{8\ell/N_{t-1}^\lambda(e)}
 \label{eq:weighted-good}
\end{equation}
simultaneously for all $t,e$ with $N_{t-1}^\lambda(e)\ge1$.
Indeed the union bound is
$2KT(1+\lceil\log_2T\rceil)(2KT^3)^{-8}\le T^{-2}$ for $K\ge1,T\ge2$.
This proof accommodates evaluation events depending on the query and on
previous graph changes. It controls a self-normalized weighted sum; an
unweighted fixed-count inequality would not suffice.

\subsection{Old evidence and low-count choices}
If the current phase starts at query $b$, the total pre-phase weight before
serving query $t$ is at most $\lambda^{t-b}/\eta$. Let
\[
 \tau_\eta=\left\lceil\frac{\log(8/(g\eta))}{\eta}\right\rceil.
\]
At most $\Upsilon_T\tau_\eta$ queries lie within this many queries of a
breakpoint. On any other query the pre-phase weight is at most $g/8$,
because $\lambda^m\le e^{-\eta m}$. There is no pre-phase weight in the
initial phase. If $N_{t-1}^\lambda(e)\ge1$, the conditional-mean average
in \eqref{eq:weighted-good} is therefore within $\kappa+g/8$ of the
current phase's reference: current-phase observations contribute bias at
most $\kappa$, and old observations contribute at most their weight divided
by $N$. We do not claim this for arbitrarily small counts.

On such a nontransient, correct-graph query satisfying
\eqref{eq:weighted-good}, let $e$ be the first wrongly selected edge and
$e^\star$ the best available edge. If $e^\star$ has discounted count at least $1$, then
$U_t^\lambda(e^\star)\ge\theta_{r,i}(e^\star)-\kappa-g/8$.
If its count is below $1$, its radius exceeds $1$ (or its index is
infinite), so this lower bound still holds. A selected edge with
\[
 N_{t-1}^\lambda(e)\ge h:=128\ell/g^2
\]
has radius at most $g/4$ and index at most
$\theta_{r,i}(e)+\kappa+5g/8$. The reference gap is at least
$2\kappa+g$, so this is strictly less than $e^\star$'s index. Thus every charged
mistake on a nontransient good query selects an edge with count below $h$.

There is a deterministic bound on these selections. Partition the horizon
into blocks of length $L_\eta=\lceil1/\eta\rceil$. Since $\lambda\ge1/2$,
$\lambda^{L_\eta}\ge1/8$. If an edge has already been evaluated $m$ times within
the current block, its discounted count is at least $m/8$. Consequently
there are at most $\lceil8h\rceil+1$ selections with count below $h$ for
that edge in the block. There are at most $T\eta+1$ blocks, giving
\begin{equation}
 \#\{\text{selected edges with }N_{t-1}^\lambda<h\}
 \le K(\lceil8h\rceil+1)(T\eta+1).
 \label{eq:discount-low-count}
\end{equation}
This bound allows sleeping edges, multiple observations on different edges
within an episode, and changing graphs. It uses neither persistent
availability nor permanent identification of an optimal arm.

\subsection{Completion of Theorem~\ref{thm:drift}}
Apply the coupling of Appendix~\ref{app:coupling} against each phase's graph
and policy. Graph-mismatch queries cost at most $HD_T^d$. Transient queries
cost at most $H\Upsilon_T\tau_\eta$. The complement of the event
\eqref{eq:weighted-good} contributes at most $HT\cdot T^{-2}=H/T$ in
expectation. All remaining first differences are charged to
\eqref{eq:discount-low-count}. Substitution gives exactly the bound in
Theorem~\ref{thm:drift}, suppressing fixed $H$ in its $O(\cdot)$ terms.
For $\eta=\min\{1/2,\sqrt{(\Upsilon_T+1)/T}\}$, the terms
$T\eta+1$ and $(\Upsilon_T+1)/\eta$ are
$O(\sqrt{(\Upsilon_T+1)T})$; the remaining factors are logarithmic for
fixed positive gaps and fixed $K$. Transients from short phases can overlap;
their union is still bounded by $\Upsilon_T\tau_\eta$, so no minimum phase
length is needed.

This argument follows the discounting principle of
\cite{garivier2011swucb}, but proves the required concentration and counting
steps for the present conditional-mean observation model. In particular,
it does not apply a running-mean stabilization proposition to an estimator
with fixed forgetting, nor infer accuracy of a ratio from small old
numerator and denominator masses.

\subsection{A sufficient bound on dynamic graph mismatch}
\label{app:dynamic-structure}
The graph term can also be controlled under stronger feedback. Suppose
every phase's target edges are available at each batch in that phase,
every compared pair is evaluated with conditional probability at least
$\nu>0$ per query, and the set margins have a common effective lower bound
$g_{\rm set}>0$. Suppose further that
\begin{equation}
 \eta\le\frac{\nu g_{\rm set}^2}{8192\ell_T}.
 \label{eq:dynamic-coverage}
\end{equation}
Then, with
$\tau_{\rm set}=\lceil\log(8/(g_{\rm set}\eta))/\eta\rceil$,
\begin{equation}
 D_T^d=O\bigl((\Upsilon_T+1)(\tau_{\rm set}+B)+1\bigr).
 \label{eq:dynamic-structure}
\end{equation}
These are sufficient conditions, not consequences of LRU candidate buffers.

To prove the claim, take the most recent $L_\eta=\lceil1/\eta\rceil$ queries
before each decision. Outside an initial prefix of length $\tau_{\rm set}$
and an equally long prefix after each change, these queries are in the
current phase. Equation~\eqref{eq:count-tail}, applied to this deterministic
interval, shows that they contain at least $\nu L_\eta/2$ observations of each
pair except with probability $e^{-\nu L_\eta/8}$. Their discount weights are
at least $1/8$, so $N^\lambda\ge\nu L_\eta/16$.
Condition~\eqref{eq:dynamic-coverage} makes this at least
$512\ell_T/g_{\rm set}^2$, and the radius in
\eqref{eq:weighted-good} is at most $g_{\rm set}/8$.
Old evidence contributes at most $g_{\rm set}/8$ by the preceding
transient calculation. Hence every score is within
$\kappa+g_{\rm set}/4$ of its current reference.
Union over all pairs and all deterministic time indices bounds the count
failures by $KT e^{-\nu L_\eta/8}\le T^{-2}$ under
\eqref{eq:dynamic-coverage}; the weighted concentration failures are
already at most $T^{-2}$. Outside these events every subsequent batch
selects the correct graph. A phase therefore incurs at most
$\tau_{\rm set}+B$ mismatched queries, and failures contribute at most
$2/T$ in expectation. The argument also applies to scores at a batch after
the current query is recorded by shifting the deterministic index by one.
This proves \eqref{eq:dynamic-structure}.

For fixed $K,\nu,g_{\rm set},g_{\rm nav},B$, the tuning of
Theorem~\ref{thm:drift} gives the full
$\widetilde O(\sqrt{(\Upsilon_T+1)T})$ dynamic-regret bound whenever
\eqref{eq:dynamic-coverage} holds. The condition explicitly states the
memory needed relative to sampling frequency and the rewiring margin.

\section{Beam Extension}
\label{app:beam}
The analyzed beam exposes all improving out-edges of the frontier whose
destinations are unvisited. It scans these proposals in score order,
accepting an edge when its destination is new and its source has
contributed fewer than $b$ accepted edges at that frontier step. The scan
stops at $W$ acceptances for the next frontier or $H$ acceptances in the
query. Thus $b$ limits accepted edges per source, whereas the experimental
variant in \S\ref{sec:beam} limits proposals before pooling.
For $\epsilon$-greedy, with probability $\epsilon_t$ the first slot samples
uniformly from all eligible proposals before the scan; UCB scans indices
directly. This lets another edge from a source compete when a preferred
destination was accepted from a different source. With $W=1$, the rule
reduces to single-path navigation.

For this rule, let $G_W^\dagger$ select the top
$\min\{M,|\mathcal U(u)|\}$ candidates at each $u$ by beam references
$\theta_{r,W}$, and $\pi_W^\dagger$ scan feasible
proposals in that order. Put $R_W^\dagger=R(G_W^\dagger,\pi_W^\dagger)$,
$\mathrm{Reg}_W(T)=TR_W^\dagger-\sum_{t=1}^T\mathbb E[R_t]$, and
$D_{T,W}=\sum_{t=1}^T\Pr(G_t\ne G_W^\dagger)$, with $G_t,R_t$ from the beam
learner and graph equality on $V_\circ$.
The formal extension uses the full-proposal feasible scan, total episode
cap $H$, and retained running-mean evidence. The observation model and
concentration arguments are unchanged, since each pair is recorded at
most once per query. No ordering of beam visitation relative to single-path
visitation is assumed.

\subsection{Reserved-slot coverage}
\label{app:lem2}
\begin{lemma}[Reserved-slot coverage]
\label{lem:beam-infinite}
Conditional on an edge being an improving, unvisited out-edge of a frontier
node, before the episode cap, its reserved-slot acceptance probability is
at least $\epsilon_t/(WM)$. Infinite evaluation follows when the resulting
conditional probabilities have divergent sum, as in Lemma~\ref{lem:infinite}.
\end{lemma}
\begin{proof}
Before any proposal is accepted at a frontier step, there are at most
$WM$ improving proposals with unvisited destination nodes. If the reserved-slot coin
fires, each has probability at least $1/(WM)$ of being chosen first. It is
then feasible because no source node has yet used its branch quota and no destination node
has yet been accepted at that step. Thus, conditional on eligibility,
the acceptance probability is at least $\epsilon_t/(WM)$. Integrating
over eligibility and applying conditional Borel--Cantelli proves the
claim whenever the cumulative conditional probabilities diverge. Sampling
only from a locally truncated proposal list would not establish this
bound for an edge omitted from that list.
\end{proof}

\subsection{UCB beam navigation}
\label{app:thm6}
\begin{theorem}[UCB beam]
\label{thm:beam-ucb}
Assume the beam rewards satisfy Assumption~\ref{ass:means}. For some
$2\kappa<\Delta_W\le1$, every pair of distinct proposals that can compete
in its feasible scan has reference separation at least $\Delta_W$.
With $g_W=\Delta_W-2\kappa$,
$|\mathrm{Reg}_W(T)|\le HD_{T,W}+O(K\log(T\vee2)/g_W^2)$.
This bounds erroneous selections, not total acceptances of every edge
below its source node's maximum.
\end{theorem}
\begin{proof}
Couple the two beams on the same query. On the correct graph, their
frontiers and accepted edges coincide up to the first difference. At its
frontier step, the accepted prefix also coincides. The remaining feasible
proposals therefore coincide: the same destination nodes are unvisited, the same
quotas at each source node remain, and the same number of frontier slots is free.
Let $e^\star$ be the highest-reference feasible proposal and $e$ the first
different proposal selected by UCB. The feasible scan ensures
$U_t(e)\ge U_t(e^\star)$, while the beam margin ensures
$\theta_{r,W}(e^\star)-\theta_{r,W}(e)\ge\Delta_W$.

The confidence event \eqref{eq:ucb-good} and the threshold argument of
Appendix~\ref{app:thm4}, with $g_W$ in place of $g_{\rm nav}$, show that
either a confidence event fails or
$N_{t-1}(e)<32\log(T\vee2)/g_W^2$. The selected proposal is accepted and
observed, so each charge increments its count. There are at most
$O(K\log(T\vee2)/g_W^2)$ charged first differences in expectation.
Collected rewards differ by at most $H$ after a difference, giving the
theorem's absolute regret bound and graph-mismatch term.

The proof compares whole feasible scans. It does not call every edge below
its source node's largest mean suboptimal: several edges from one source node can correctly
occupy different acceptance slots. Nor does it assume a cardinality-only
multiple-play problem when quotas at source nodes and duplicate destination nodes impose
additional constraints. The same first-difference comparison, together
with \eqref{eq:weighted-good} and \eqref{eq:discount-low-count}, proves the
discounted extension claimed in Remark~\ref{rem:beam-carry}.
\end{proof}

\begin{remark}
\label{rem:beam-carry}
Estimation and stabilization apply to the beam under their stated observation
and availability conditions. The discounted-UCB proof also applies to this
feasible scan using $g_W$ and the same total episode cap $H$. Width alone
does not establish coverage, a navigation gap, or a beam-versus-path
advantage; these are not consequences of multiplying constants by $W$.
\end{remark}

\section{Self-Supervised and Mixed Feedback}
\label{app:selfsup}
\subsection{Aggregate endpoint-label bias}
\label{app:lem3}
Use the examined-node set $\mathcal X_{\rm eval}(q)$ defined in
\S\ref{sec:algorithm}, and assume at least $k$ nodes are examined.
Self-labels are the returned
set $A(q)=\operatorname{top}_k\mathcal X_{\rm eval}(q)$, using consistent
distance ties. For a fixed graph $G$ and policy $\pi$, let
$\widetilde R(G,\pi)$ be the expected total endpoint reward using these
self-labels, and let $R(G,\pi)$ use exact labels on the same search.
Write $g(G,\pi)=\mathbb E[1-\mathrm{Recall}_k(q)]$ for the expected
fraction of true neighbors missed.

\begin{lemma}[Self-feedback bias]
\label{lem:bias}
For the fixed graph, policy, and common search above,
$|\widetilde R(G,\pi)-R(G,\pi)|\le k\,g(G,\pi)$.
\end{lemma}
\begin{proof}
Let $A=A(q)$, $B=N_k(q)$, and $S$ be the set of accepted destination nodes.
Since destination nodes do not repeat, $\widetilde\rho(q)=|S\cap A|$ and
$\rho(q)=|S\cap B|$. Their difference is
$|S\cap(A\setminus B)|-|S\cap(B\setminus A)|$.
Both counts lie in $[0,k-|A\cap B|]$, because $|A|=|B|=k$. Thus
\[
 |\widetilde\rho(q)-\rho(q)|\le k-|A\cap B|
 =k(1-\mathrm{Recall}_k(q)).
\]
Taking expectations and applying $|\mathbb E Z|\le\mathbb E|Z|$ proves the
bound. If the two feedback sources change the current trajectory, this
comparison does not apply. The bound is aggregate; conditioning on a rare
edge evaluation can magnify bias, so it does not imply a bound on
$|\theta_{\rm self}(e)-\theta_{\rm ex}(e)|$.
\end{proof}

\subsection{Mixed-feedback estimation and graph recovery}
\label{app:thm7}
\begin{theorem}[Mixed-feedback estimation and graph recovery]
\label{thm:tworate}
Let $r_{\rm ex}$ and $r_{\rm self}$ be the rewards the two feedback modes
would assign to an evaluated edge on the same completed search, both
bounded by $R_{\max}$ and nonnegative. Select exact feedback with
probability $p$, independently of the history and current-query randomness,
while keeping the search and evaluated edges fixed. For each mode
$j\in\{{\rm ex},{\rm self}\}$, suppose the mean reward, conditional on
the prior history and evaluation of edge $e$, stays within $\kappa_j$ of
a fixed reference value $\theta_j(e)$, the mode-specific counterpart of
$\theta_r(e)$ in Assumption~\ref{ass:means}. Retain accumulated reward sums and
observation counts across rewiring and eviction, and use running-mean
scores. Define the mixed reference and tolerance
\[
 \theta_p=p\theta_{\rm ex}+(1-p)\theta_{\rm self},\qquad
 \kappa_p=p\kappa_{\rm ex}+(1-p)\kappa_{\rm self}.
\]
Every infinitely evaluated edge satisfies
$\limsup_t|\widehat Q_t(e)-\theta_p(e)|\le\kappa_p$ almost surely.
Over the fixed finite comparison universes of \S\ref{sec:pool}, suppose
every compared edge is evaluated infinitely often, the top-$M$ edges by
$\theta_p$ are available at every sufficiently late batch, and their set margin
$\Delta_{p,{\rm set}}>2\kappa_p$. This margin is the minimum gap between
selected and excluded reference values over nodes with a selection boundary.
Then the graph stabilizes almost surely at $G_p^\dagger$, the graph
selecting those edges.

For exact-reference recovery, let
$b_{\rm label}=\sup_e|\theta_{\rm self}(e)-\theta_{\rm ex}(e)|$, and let
$\Delta_{{\rm ex},{\rm set}}$ be the corresponding set margin under
$\theta_{\rm ex}$. Under the same estimation hypotheses, suppose every compared
edge is evaluated infinitely often and the top-$M$ edges by $\theta_{\rm ex}$
are available at every sufficiently late batch. If
\eqref{eq:tworate-exact-margin} holds, the graph stabilizes almost surely
at the exact-reference graph selecting these edges. These replace the
mixed-reference margin and availability requirements.

Expected additional feedback cost per query is
$pc_{\rm ex}+(1-p)c_{\rm self}$, where $c_{\rm ex}$ and $c_{\rm self}$
are the respective expected feedback-mode costs, including any label search.
\end{theorem}
\begin{proof}
Fix an evaluated pair and let $Z_t$ be the exact-feedback flag. By the
stated independence and common evaluation event,
\[
 \begin{gathered}
 Y_t=Z_t r_{{\rm ex},t}+(1-Z_t)r_{{\rm self},t},\\
 \mathbb E[Y_t\mid\mathcal F_{t-1},I_t=1]
 =p\mu_{{\rm ex},t}+(1-p)\mu_{{\rm self},t}.
 \end{gathered}
\]
The latter mean lies within $\kappa_p$ of $\theta_p$. Scaling the common
concentration argument by $R_{\max}$ proves the almost-sure estimation
claim. Let $\tau_m(e)$ be the $m$th observation time and $\widehat Q^{[m]}(e)$
its score. For fixed $m\ge1$ and $\delta\in(0,1)$, \eqref{eq:sample-tail} gives
\[
 \Pr\!\left(
 \begin{gathered}
 \tau_m(e)<\infty,\\
 |\widehat Q^{[m]}(e)-\theta_p(e)|>
 \kappa_p+R_{\max}\sqrt{\log(2/\delta)/(2m)}
 \end{gathered}
 \right)\le\delta.
\]
The stabilization argument of Appendix~\ref{app:prop1} applies with
$\theta_p,\kappa_p,\Delta_{p,{\rm set}}$ in place of the corresponding single-source
quantities. No independence between different edges of the same query is
required. Evidence must not be reset at each re-entry.

For agreement with the exact reference, the triangle inequality gives
\[
 \limsup_t|\widehat Q_t(e)-\theta_{\rm ex}(e)|
 \le\kappa_p+(1-p)b_{\rm label}.
\]
Condition~\eqref{eq:tworate-exact-margin} separates every exact-reference
selected edge from every excluded edge in empirical score, once estimates
are sufficiently accurate. Infinite evaluation of every compared pair
makes these inequalities eventual simultaneously. Availability of the
exact-reference edges at every sufficiently late batch then gives the
claimed graph. Comparison with a separate $p=1$ learner
requires common references and that learner's own hypotheses; different
proposal histories can otherwise induce different conditional values.

Finally, conditioning on the feedback branch gives expected cost
$pc_{\rm ex}+(1-p)c_{\rm self}$ by total expectation. This includes any
additional teacher search. A $p$-thinning of exact flags does not by itself
give a graph-learning time proportional to $1/p$, because candidate
availability, evaluation frequencies, and the remaining margin also matter.
\end{proof}

The theorem covers endpoint and recall rewards (\S\ref{sec:reward}) and
bounded MC returns (\S\ref{sec:gammavariants}), under its stated hypotheses;
endpoint MC($0.8$) permits $R_{\max}=5$.

\fi

\end{document}